\documentclass[%
 superscriptaddress,
 nofootinbib,
 amsmath,amssymb,
 aps, prd,
 floatfix
]{revtex4-2}
\usepackage{booktabs}
\usepackage{hyperref}
\usepackage{bm}
\usepackage{subcaption}
\usepackage[multidot]{grffile}
\usepackage{soul}
\usepackage[dvipsnames]{xcolor}
\newcommand{\bea}{\begin{eqnarray}}
\newcommand{\eea}{\end{eqnarray}}

\newcommand{\Cath}{\text{Ca}_\text{th}}
\newcommand{\Ca}{\text{Ca}}

\usepackage{lineno}

\usepackage[utf8]{inputenc}

\newcounter{Cequ}
\newenvironment{CEquation}
  {\stepcounter{Cequ}%
    \addtocounter{equation}{-1}%
    \equation}
  {\endequation}

\begin{document}

\title[Astrocyte-Mediated Higher-Order Control of Synaptic Plasticity]{Astrocyte-Mediated Higher-Order Control of Synaptic Plasticity}
\author{Gustavo Menesse}
\affiliation{Department of Electromagnetism and Matter Physics, and Institute ``Carlos I'' of Theoretical and Computational Physics, University of Granada, Granada, Spain} 
\author{Ana P. Mill\'an}
\affiliation{Department of Electromagnetism and Matter Physics, and Institute ``Carlos I'' of Theoretical and Computational Physics, University of Granada, Granada, Spain} 
\author{Joaqu\'{\i}n J. Torres}
\email[]{jtorres@onsager.ugr.es: Corresponding author}
\affiliation{Department of Electromagnetism and Matter Physics, and Institute ``Carlos I'' of Theoretical and Computational Physics, University of Granada, Granada, Spain}

\begin{abstract}
The dynamics of higher-order topological signals are increasingly recognized as a key aspect of the {activity} of complex systems. 
A paradigmatic example are synaptic dynamics: synaptic efficacy changes over time {driven} by different mechanisms. 
Beyond traditional node-driven short-term plasticity mechanisms, the role of astrocyte modulation through higher-order {interactions}, in the so-called tripartite synapse, is increasingly recognized. 
However, the competition and interplay between node-driven and higher-order mechanisms {have} yet to be considered. 
Here, we introduce a simple higher-order {model of the tripartite synapse accounting for astrocyte-synapse-neuron interactions in short-term plasticity. In the model,}  
%astrocyte-neuron short-term plasticity model in which 
astrocyte gliotransmission and pre-synaptic intrinsic facilitation mechanisms jointly modulate the probability of neurotransmitter release {at the synapse}, generalizing previous short-term plasticity models.
We investigate the implications of such mechanisms in a minimal recurrent motif -- a directed ring of three excitatory leaky integrate-and-fire neurons -- where one neuron receives external stimulation that propagates through the circuit. 
Due to its strong recurrence, the circuit is highly prone to self-sustained activity, which can make it insensitive to external input. 
By introducing higher-order interactions {among} different synapses through astrocyte modulation, we show that {higher}-order modulation robustly stabilizes circuit dynamics and expands the parameter space that supports stimulus-driven activity. 
Our findings highlight a plausible mechanism by which astrocytes can reshape effective connectivity and enhance information processing through {higher}-order structural interactions -- even in the simplest recurrent circuits.
\end{abstract}

\maketitle

\section{Introduction}
Synaptic interactions between neurons form a dynamic, adaptive network that shapes brain dynamics and emergent cognitive phenomena. 
Such structure–function coupling is a defining feature of complex systems in which underlying connectivity shapes -- and is shaped by -- emergent behavior. 
Typically the system dynamics are associated only with the nodes of the network, which in the case of neuronal networks stand for neurons. 
Recently however it has become evident that the edges between the nodes -- the synapses -- can also be associated with topological signals, i.e. dynamical variables which are not necessarily reduced to node dynamics \cite{naturephysicsana2025}. 
Topological signals are ubiquitous {in complex systems}, and include molecular transportation fluxes in cells, synaptic signals in the brain, dynamic functional connectivity among brain regions \cite{faskowitz2022edges, santoro2023higher, santoro2024higher, santos2023emergence, tewarie2023non}, or vector fields such as the one representing currents in the ocean \cite{schaub2020random}. 
Each topological signal is associated with a dimension {that emerges from the underlying topological substrate of simplicial and cell complexes \cite{barbarossa2020topological} and} reflects the type of interaction {the signal} represents: 
signals on nodes are $0$-dimensional, those on edges are $1$-dimensional, those on filled triangles or cycles (forming polygons) are $2$-dimensional and so on. 
{Simplicial complexes and topological signals }
provide a natural way to model and analyze complex, multi-scale interactions within a unified mathematical framework \cite{barbarossa2020topological}. 
{The emerging research field of higher-order topological dynamics \cite{naturephysicsana2025}, combining the study of non-linear dynamics with algebraic topology to model and study the dynamics of higher-order topological signals, has revealed the emergence of new dynamical states in such systems \cite{battiston2020, millan2025, digaetano2024neighbourhood, stolz2021, Torres21, cencetti2023distinguishing}.
}
{The impact of higher-order topological dynamics goes beyond the study of complex systems. Recently, for instance, it has} 
led to a new general framework for AI algorithms that can take advantage of the novel dynamics that are uniquely possible for higher-order topological signals \cite{zhou2006learning, sardellitti2024topological, zhao2023dynamic, ji2022fc, antelmi2023survey}. 

Synaptic signals are a paradigmatic example of higher-order topological signals. {Synapses are typically modeled as the ($1$-dimensional) edges of neuronal networks and, notably, their efficacy changes} 
dynamically over time both in short and long time-scales via \emph{synaptic plasticity} mechanisms. 
Among these, short-term plasticity (STP) plays a crucial role in modulating the on-line transmission of the neural code by dynamically adjusting synaptic strength on timescales of milliseconds to seconds \cite{Abbott2004}. STP includes transient facilitation (STF) and depression (STD), which depend primarily on the recent activity of the presynaptic neuron and govern the immediate response of a synapse to successive stimuli \cite{STEVENS1995, Markram1996, Lin2000, Li2003}. This rapid, activity-dependent modulation acts as a dynamic filter that can emphasize or suppress specific temporal patterns of neural activity, effectively shaping information flow in real time. A well-established model of STP introduced in Ref. \cite{Tsodyks1998} captures both facilitation and depression, and has been shown to generate non-equilibrium phases linked to dynamic memory and resonance phenomena near criticality \cite{torreskappen2013, torresmarro2015, biologypretel, menesse2024}. In contrast to long-term plasticity, which involves lasting changes driven by correlated pre- and post-synaptic activity, STP provides a fast, flexible mechanism for regulating synaptic efficacy and supporting context-dependent computation in neural circuits.

Over the past two decades, accumulating evidence has shown that synaptic plasticity involves not only pair-wise, neuron-to-neuron communication, but also mechanisms that are intrinsically higher-order. A paradigmatic example is the tripartite synapse, in which astrocytes—non-neuronal glial cells and the most abundant cell type in the brain—play an active role in modulating synaptic transmission \cite{Araque1999, DePitta2016}. Unlike traditional neuron-centric models, where synaptic plasticity is viewed as a local, node-driven process, the tripartite model reveals a more distributed and collective form of regulation. Astrocytes can simultaneously monitor and modulate the activity of multiple neighboring synapses, often across different neurons, exerting control over synaptic efficacy and, consequently, over information flow in neural circuits \cite{Bazargani2016, DePitta2019}.
This glial modulation is not merely auxiliary but may constitute a core mechanism of synaptic computation, contributing to the temporal filtering, integration, and stabilization of synaptic signals \cite{DePitta2016, DePitta2019}. 
The effects of astrocyte modulation compete with those of traditional node-driven STP, a fact that is typically {overlooked} both in STP and triparte-synapse models. 
From a higher-order perspective, however, such effects arise naturally \cite{naturephysicsana2025}. In this context, a neuronal network made of neurons and synapses is expanded by coupling the synapses through a new, higher-dimensional element, the astrocytes. 
Topologically, astrocytes are thus modeled as the $2$-dimensional polygons of the higher-order network (or cell complex), in which nodes and synapses are respectively the $0$-dimensional nodes and the $1$-dimensional edges. 
And in a higher-order network modeling the dynamics of edge-signal (synaptic efficacies) naturally involves interactions through the nodes (neurons) they share, as well as through the polygons (the astrocytes) they conform \cite{millan2020explosive, schaub2020random}. 

Another example of higher-order modulation is the axo-axonic synapse, where a neuron directly targets the axon of another neuron to modulate the efficacy of its downstream synaptic connections—typically exerting an inhibitory effect \cite{cover2021axo, pan2020activity}. Recent work has framed axo-axonic interactions as a form of triadic control over edge dynamics, leading to emergent spatiotemporal patterns in recurrent neural circuits \cite{millan2025, sun2023dynamic}.
Together, these mechanisms support the view that synaptic computation is not a purely dyadic phenomenon but a multi-agent, higher-order process, with astrocytes playing a particularly central role in the orchestration of dynamic and context-sensitive plasticity.

From the neuroscience perspective, modeling of astrocyte signaling is a significant challenge due to their complex morphology and diverse dynamics. 
Astrocytes regulate synaptic plasticity by different mechanisms, including gliotransmission \cite{Ceglia2023}, direct regulation of extracellular ion concentrations, and neurotransmitter uptake that clears the synaptic cleft \cite{Lenk2024}.
In particular, gliotransmission is a main process in neuron-astrocyte interaction by which astrocytes (and more generally glia cells) release gliotransmitters such as glutamate (a key excitatory neurotransmitter) that interact directly with different receptors of both the pre- and post-synaptic neurons \cite{Ceglia2023}.
Among other effects \cite{Kovacs2017, Csemer2023, Jo2014} gliotransmission plays a role both in short- \cite{Pitta2022} and long-term \cite{Durkee2021} plasticity by modulating the probability of neurotransmitter release at the pre-synaptic neuron, either increasing or decreasing it \cite{Navarrete2008, DePitta2011}. 
Models of astrocyte regulation of different levels of detail have been proposed \cite{DePitta2012, Oschmann2018, Lenk2024}. 
Many of these extend traditional short- and long-term synaptic dynamics to incorporate astrocyte influences, ranging from biophysically detailed compartmental models \cite{Manninen2020, Verisokin2021} to more simplified descriptions \cite{Lenk2020}.
Overall, previous models have neglected the interplay between pre-synaptic and astrocyte mechanisms. 
Given than both of them act on the same neurotransmitters, exploring this competition remains an open question with potentially far-reaching implications.

In order to address this gap, in this study we propose a simple higher-order astrocyte-{synapse-}neuron short-term plasticity model ({ASN}-STP) that captures the competition between astrocyte 
and pre-synaptic mechanisms in short-term facilitation dynamics. 
As we go on to show, our general framework reduces to previous models in the literature \cite{Tsodyks1998, Pitta2022} for specific simplifications. 
By examining the dynamics of the tripartite synapse from a higher-order perspective \cite{millan2025}, our model provides a unified framework to study the interplay between neuronal (dimension zero), synaptic (dimension one) and astrocyte (dimension two) dynamics, and how these shape emergent neuronal activity. 

To illustrate the emergent dynamics of the system, we consider in particular the role of astrocyte regulation in a simple ring of neurons connected by excitatory synapses. 
Our findings illustrate that the presence of even a single tripartite synapse is sufficient to prevent runaway excitatory activity. This effect is strongest when higher-order interactions are present, such that a single astrocyte modulates several synapses coordinately, as it is the case in brain networks. 
Higher-order interactions further enhance the system’s responsiveness to external stimuli when they act on internal synapses, as opposed to those responding directly external stimuli, suggesting that astrocyte modulation plays a more central role in deeper, integrative circuits, rather than sensory ones. The fact that tripartite synapse are more abundant in the hippocampus and cerebellum \cite{Tan2021}, present in more than $50 \%$ of all synapses in the hippocampus \cite{Ventura1999}, which are regions with high internal recurrence \cite{LeDuigou2014}\cite{Sammons2024} support the biological plausibility of our theoretical results. In resume, higher-order astrocyte modulation is highly beneficial to modulate plasticity and neuronal activity of high-recurrent circuits.

%%%%%%%%%%%%%%%%%%%%%%%%%%%%%%%%%%%%%%%%%%%%%%%%%%%%%%%%%%%%%%%%%%%

\section{Higher-order neuron-synapse-astrocyte model}
\begin{figure}[htb]
\centering
\includegraphics[width=\linewidth]{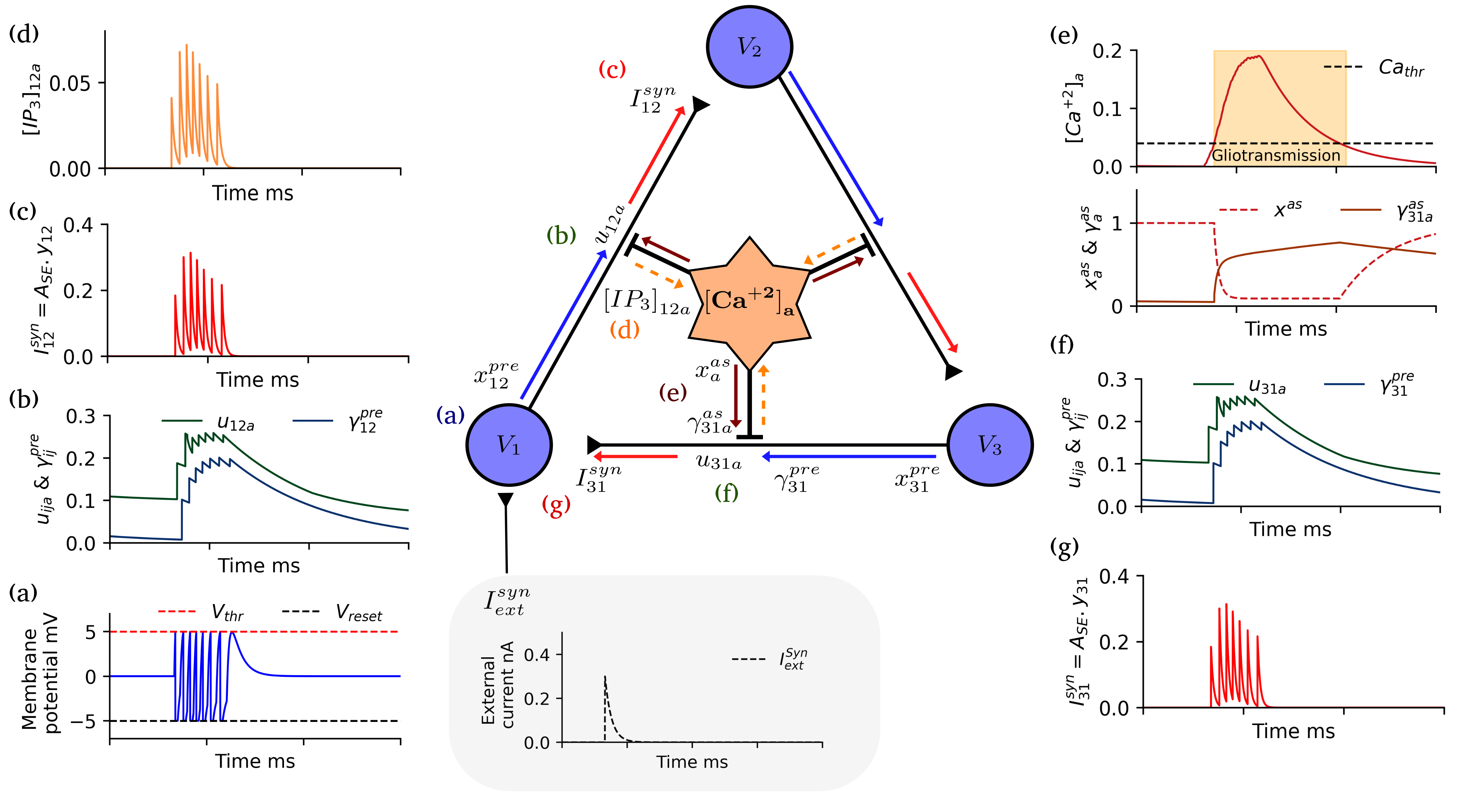}
\caption{
\textbf{Model scheme in a minimal recurrent excitatory circuit.}
We consider a minimal recurrent network consisting of three excitatory neurons arranged in a ring, with states $V_i$.
An external input $I^{syn}_{ext}$ arrives at neuron $1$ inducing spiking activity, as shown in panel (a), which propagates through the circuit. 
Panel (b) show the synaptic variables $u_{12}^a$ (neurotransmitter release probability) and $\gamma_{12,\text{pre}}^a$ (pre-synaptic facilitation variable), which increase following pre-synaptic activity. 
Panel (c) shows the amplitude of the synaptic currents at synapse $12$, namely $A_{SE}y_{12}$ (eq. \ref{eq:I_syn}), which is affected both by the dynamics of neuron $1$ and the astrocyte $a$.
Panel (d) shows the IP$_3$ dynamics in the astrocyte process at synapse $ij$, displaying the increase in $[IP_3]_{12}^a$ following pre-synaptic spikes. 
Panel (e) shows the dynamics of the astrocyte variables: the calcium concentration $[\Ca^{+2}]^a$ (top panel), and the astrocyte depletion and facilitation variables $x^a_\text{astro}$ and $\gamma_{ij, \text{astro}}^a$.  \textbf{High-order interaction} By astrocyte mediation, the activity on synapse $12$, affects facilitation in synapse $31$ (Panel f), which change the recurrence current in Neuron $1$ (Panel g). 
}
\label{fig:Exp_scheme}
\end{figure}

Here we propose a simple astrocyte{-synapse}-neuron short-term plasticity model ({ASN}-STP) that captures the potential competition between gliotransmission, mediated by astrocytes, and pre-synaptic mechanisms, mediated by the pre-synaptic neuron, as short-term facilitation.
Our model includes the description of neuronal, synaptic, and astrocyte dynamics, as well as the interactions among them. 
Neurons constitute the nodes of the higher-order network, and thus we refer to their dynamics as dimension-$0$. 
Synapses connect neurons in a directed manner, with the pre-synaptic neuron projecting onto the post-synaptic neuron. 
Synaptic dynamics are represented as $1$-dimensional topological signals taking place on the edges of the higher-order network. 
Finally, astrocytes modulate synaptic dynamics in an aggregative, non-linear manner. 
In actual brains, one astrocyte modulates several synapses at the same time, whereas each synapse can have at most one adjacent astrocyte as astrocyte will usually have non-overlapping territories \cite{Allen2017}. 
Given that astrocytes couple the $1$-dimensional edge signals, they are represented as $2$-dimensional topological signals. 
An exemplary minimal circuit composed by three neurons arranged in a loop, with three directed excitatory synapses, and one astrocyte modulating the three synapses, is shown in Figure \ref{fig:Exp_scheme}. 

Neuronal, synaptic, and astrocyte dynamics used in the present work are defined in detail in the following subsections. 
Here we describe them in a summary manner. {The reference values of all parameters are detailed in Table \ref{tab:Parameters}.}
Neurons are defined as leaky integrate-and-fire units whose state is given by a voltage variable $V_i(t)$ (eq. \ref{Eq_LIF}), such that the neuron is said to fire or release an action potential or spike when the voltage reaches a threshold. 
Afterwards, $V_i(t)$ is reset to a hyperpolarized (negative) value. 
Synapses are characterized by the fraction of active neurotransmitters $y_{ij}$ (eq. \ref{eq:yij_neuron_synapse}). 
Each time the pre-synaptic neuron fires, $y_{ij}$ increases depending on three factors: 
the fraction $\alpha$ of neurotransmitters that remain at the synapse, the probability of neurotransmitter release $u_{ij}$, 
and the fraction of available neurotransmitters $x_{ij}$.
The parameter $\alpha$ accounts for the competition between the astrocyte and the synapse for released neurotransmitters. 
A fraction $1-\alpha$ of released neurotransmitter is recruited by the astrocyte {(eq. \ref{eq:1-alpha})}, leaving only a fraction $\alpha$ available for synaptic transmision. 
Thus $\alpha$ is a main control parameter of the dynamics.
The synaptic variables $u_{ij}$ (eq. \ref{Eq_U_fin})  and $x_{ij}$ {(eq. \ref{eq:depletion})} account respectively for short-term facilitation and depression mechanisms.

STF is affected both by presynaptic mechanisms driven by the pre-synaptic neuron and gliotransmission driven by the astrocyte. 
Here, we consider both effects, as well as the competition between them. 
The neuron-driven plasticity dynamics is inspired by previous models of neuron-driven STP \cite{Markram1996, Tsodyks1998}. 
Regarding astrocytes, they affect STF through gliotransmission, the process by which they recruit and release gliotransmitters at each {adjacent} synapse \cite{Pitta2022}. 
Astrocyte dynamics is usually described by variations of its cytosolic calcium concentration, $[\Ca^{2+}]^a$. 
{$[\Ca^{2+}]^a$ increases following each pre-synaptic neuron spike at any synapse regulated by the astrocyte, from which
the astrocyte recruits a fraction of the released neurotransmitters proportional to $1-\alpha$}.  
When a {$[\Ca^{2+}]^a$} threshold is reached, we assume the astrocyte to become active and release gliotransmitters into its {adjacent} synapses {$ij$}, modulating {their} pre-synaptic release probability $u_{ij}$ (eq. \ref{Eq_U_fin}). 
Contrary to neuron-driven STF, gliotransmission can lead to both an increase and a decrease of the probability of neurotransmitter release depending on a control parameter {(namely $\varepsilon$ in equation \ref{Eq_U_fin})} \cite{DePitt2011}.
Moreover, recruitment of neurotransmitters by the astrocyte competes with neuron-driven STF, naturally leading to competition effects that cannot be recovered by the independent original models, as we go on to show. 
{The competition is affected by the relative temporal scales of each mechanisms. Pre-synaptic STF occurs on scales of hundreds of milliseconds to a few seconds, whereas gliotransmission is slower and takes place in the time-scale of seconds to minutes. This fact, and its consequences, is directly captured with our proposed model.}
Finally, we also consider short-term depression in the model, reflecting the depletion of neurotransmitters after repeated resources release, and which occurs both at the synapse and astrocyte levels (eqs. \ref{eq:depletion}). 

\subsection{Dimension zero dynamics: the neurons}

The neuronal population is modeled using the leaky integrate-and-fire (LIF) model. 
If the membrane potential of the neuron is below a threshold $V_{th}$, the dynamics follow:
\begin{equation}
    \tau_{V}\frac{dV_{j}}{dt} = - V_{j} + R I_{j}^{stim}(t) + R I_{j}^{syn}(t) \:, \label{Eq_LIF}
\end{equation}
where $V_j$ is the membrane potential of neuron $j$, $R$ the membrane resistance, and $\tau_V$ the membrane time constant.
When the membrane potential crosses the threshold, i.e., $V_j(t) \geq V_{th}$ at time $t = t_j^{spk}$, a spike is emitted by neuron $j$, and the membrane potential is reset to a hyperpolarized value, $V(t_j^{spk}) = V_{reset}$ mV. 
Following this, the neuron enters an absolute refractory period $t_a$, during which the potential is clamped at $V_{reset}$.

{An input current $I_{1}^{stim}(t)$ arrives at neuron $1$ ($I_{j\neq 1}^{stim}(t)=0$), which} is modeled as a series of exponentially decaying pulses applied at specific times $t_k^{stim}$, {namely}
\begin{equation}
    I_{1}^{stim}(t) = \sum_{k} A_{ext} \exp\left( \frac{t_k^{stim} - t}{\tau_{in}} \right) \Theta(t - t_k^{stim}) \:,
\end{equation}
where $A_{ext}$ is the stimulus amplitude, $\tau_{in}$ the decay time constant, and $\Theta(\cdot)$ the Heaviside step function.
The synaptic current received by neuron $j$ is, including short-term plasticity mechanisms:
\begin{equation}\label{eq:I_syn}
    I_j^{syn}(t) = \sum_{i \in nn(j)} \mathcal{A}_{SE} \, y_{ij}(t) \:,
\end{equation}
where $nn(j)$ denotes the set of pre-synaptic neurons projecting onto neuron $j$, and $\mathcal{A}_{SE}$ {is} the absolute synaptic efficacy.
The variable $y_{ij}(t)$ indicates the synaptic efficacy: it stands for fraction of active neurotransmitters in synapse $ij$, i.e., the amount of neurotransmitters that bound to post-synaptic receptors. 
The dynamics of $y_{ij}(t)$ depend both on node- and astrocyte-driven short-term mechanisms, as we describe in the following subsections.

\subsection{Dimension one dynamics: the synapses}
Lets consider first that {each synapse is modulated by one astrocyte. In the following we indicate this in all synaptic variables by the super-index $a$, indicating the astrocyte that modulates the synapse.}
The fraction of active neurotransmitter $y^a_{ij}(t)$ accounting for the dynamical state of synapse $ij$, {modulated by astrocyte $a$,} evolves according to (see \citep{Tsodyks1998}):
\begin{equation}\label{eq:yij_neuron_synapse}
    \frac{dy_{ij}^a}{dt} = -\frac{y^a_{ij}}{\tau_{in}} + \alpha \, x^a_{ij}(t) \, u^a_{ij}(t) \, \delta(t - t_i^{spk}) \:,
\end{equation}
Here $\alpha$ is a main control parameter of the model, representing  the fraction of neurotransmitters that remain in the synaptic cleft and contribute to the post-synaptic current (eq. \ref{eq:I_syn}).
A fraction $1-\alpha$ of neurotransmitter escapes the synaptic cleft and activates astrocyte receptors instead of post-synaptic ones. 
The explicit inclusion of $\alpha$ thus allows us to differentiate between synaptic and astrocyte activation.
The synaptic variables $x^a_{ij}(t)$ and $u^a_{ij}(t)$ account respectively for the fraction of available synaptic resources (neurotransmitters) and for the neurotransmitter release probability at synapse $ij$, and account for STP mechanisms.  
In the following we discuss firstly facilitation and secondly depression mechanisms.

\subsubsection{Short-term facilitation: node- and astrocyte-driven mechanisms}

To model STF when both pre-synaptic and gliotransmission mechanisms are present, we extend the strategy presented in Ref. \cite{Pitta2022}. 
Firstly, we hypothesize that $u_{ij}^a(t)$ depends on the fraction of receptors and channels activated by a) gliotransmission and b) pre-synaptic activity, namely $\gamma_{ij,\text{astro}}^{a}(t)$ and $\gamma_{ij,\text{pre}}^a(t)$.
Secondly, in the absence of neuronal of astrocyte inputs, $u_{ij}^a(t)$ should reach a steady-state $U_{SE}$ (where $u$ stands for ``utilization'' and $SE$ for ``synaptic efficacy''), as originally introduced in the Tsodyks-Markram model for STP \cite{Markram1996, Tsodyks1998}.
We further hypothesize that, near its steady state, $u_{ij}^a(t)$ is a smooth function that we can approximate by its first-order power expansion, i.e.,  
\begin{eqnarray}
u_{ij}^a(\gamma_{ij,\text{pre}}^{a},\gamma_{ij,\text{astro}}^a,t) &\approx& 
U_{SE} + 
\left. \frac{\partial u_{ij}^a}{\partial \gamma_{ij,\text{astro}}^{a}}\right|_{(0,0)} \gamma_{ij,\text{astro}}^{a}(t) + 
\left. \frac{\partial u_{ij}^a}{\partial \gamma_{ij,\text{pre}}^{a}}\right|_{(0,0)} \gamma_{ij,\text{pre}}^{a}(t) \:.
\label{Eq_Utay}    
\end{eqnarray}
An important detail is the timescale difference between the two mechanisms: pre-synaptic facilitation typically operates over hundreds of milliseconds to a few seconds, whereas gliotransmission {can} influence synaptic function over longer timescales (seconds to minutes) \cite{DePitta2019, Pitta2022}. 
This justifies modeling both activated fractions with separate dynamics and distinct time constants.

The fractions of active neurotransmitters are bounded between $0$ and $1$, as well as the total fraction, i.e. $0 \leq \gamma_{ij,\text{pre}}^{a}(t) + \gamma_{ij,\text{astro}}^{a}(t) \leq 1$.
Moreover, given that $u_{ij}^a$ is a probability, it is also bounded between $0$ and $1$, and the derivatives {in eq. \ref{Eq_Utay}} must be defined to maintain these bounds. 
In particular, for the pre-synaptic mechanism we set 
\begin{equation}
\left.\frac{\partial u_{ij}^a}{\partial \gamma_{ij,\text{pre}}^{a}}\right|_{(0,0)} = 1 - U_{SE} \:,
\end{equation}
which recovers the Tsodyks-Markram dynamics in the absence of gliotransmission. 
Regarding gliotransmission, we follow the reasoning in Ref. \cite{DePitta2019} and define
\begin{equation}
  \left.\frac{\partial u_{ij}^a}{\partial \gamma_{ij,\text{astro}}^{a}}\right|_{(0,0)} = \varepsilon - U_{SE} \:,  
\end{equation}
where $0 < \varepsilon < 1$. 
This allows us to model the dual nature of gliotransmission, i.e. its capacity to either promote or inhibit neurotransmitter release. 
{In particular,}
if $\varepsilon > U_{SE}$, gliotransmission increases the release probability; otherwise, it decreases it.
We note that these definitions ensure that $u_{ij}^a$ is bounded between $0$ and $1$ both in the case of no transmission (in which case  $\gamma_{ij, \text{astro}}^{a} = \gamma_{ij,\text{pre}}^{a} = 0$ and $u_{ij}^a = U_{SE}$), and maximum transmission ($\gamma_{ij}^{pre} + \gamma_{ij, \text{astro}}^{a}=0$).

Next, we define the dynamics of the fractions of activated neurotransmitters. 
Assuming that the onset of receptor activation's effect on synaptic release is much faster than its decay time for both mechanisms, the coupled dynamics are described as:
\begin{eqnarray}
\frac{d\gamma_{ij,\text{astro}}^{a}}{dt} &=& 
-\frac{\gamma_{ij,\text{astro}}^{a}}{\tau_{f,\text{astro}}} 
+ G^{a}(t)\left[1 - (\gamma_{ij,\text{astro}}^{a} 
+ \gamma_{ij,\text{pre}}^{a})\right] \Theta([\Ca^{2+}]^{a}(t) - \Cath) \:, \\
\frac{d\gamma_{ij,\text{pre}}^{a}}{dt} &=& 
-\frac{\gamma_{ij,\text{pre}}^{a}}{\tau_{f,\text{pre}}} 
+ U_{SE}\left[1 - (\gamma_{ij,\text{astro}}^{a} 
+ \gamma_{ij,\text{pre}}^{a})\right]\delta(t - t^{sp}_{i}) \:, \label{Eq_preact}
\end{eqnarray}
where $G^{a}(t)$ is the gliotransmitter release efficacy, {which is proportional to} the availability of gliotransmitters at the astrocyte \cite{DePitta2011}, $x^{a}_{\text{astro}}(t)$,  namely
\begin{eqnarray}
G^{a}(t) = x^{a}_{\text{astro}}(t) U_{\text{astro}},
\end{eqnarray}
{where $U_{\text{astro}}$ is a constant setting the release probability.}
Gliotransmitter release is assumed to be a continuous process occurring as long as the astrocyte calcium concentration remains above the threshold $\Cath$ and resources are available. 
In contrast, the pre-synaptic facilitation mechanism -- driven by voltage-dependent calcium channels -- is triggered directly by the arrival of an action potential and is independent on the amount of neurotransmitter concentration in the synaptic cleft. 
Thus, we use the same efficacy $U_{SE}$, which ensures that in the absence of gliotransmission, the mechanism recovers the original facilitation model proposed by \cite{Markram1996,Tsodyks1998}.

The release probability is ultimately expressed as 
\begin{equation}
u_{ij}^a(\gamma_{ij,\text{astro}}^{a}, \gamma_{ij,\text{pre}}^{a},t) = U_{SE} + (\varepsilon - U_{SE})\gamma_{ij,\text{astro}}^{a}(t) + (1 - U_{SE})\gamma_{ij,\text{pre}}^{a}(t) \:, \label{Eq_U_fin}
\end{equation}
where one needs to solve the coupled dynamics of $\gamma_{ij,\text{pre}}^{a}(t)$ and $\gamma_{ij,\text{astro}}^{a}(t)$.
Our proposed facilitation model generalizes previous models as we show in the Supplementary Information (Section $1$). {In particular}, we demonstrate how our model reduces to the classical facilitation model proposed in \cite{Markram1996,Tsodyks1998} in the absence of gliotransmission, and how it becomes equivalent to the model in \cite{Pitta2022} in the absence of pre-synaptic facilitation.

\subsubsection{Resource depletion}
Due to resource depletion, neither neurotransmission nor gliotransmission can occur indefinitely under repeated stimulation. 
To model this, we consider a depression mechanism such that the fraction of available neurotransmitters and gliotransmitters decreases with each release event. 
{The former} is implemented via a short-term depression model for the pre-synaptic spine \cite{Tsodyks1998}, {and we introduce a novel} analogous mechanism for {the latter}:
\begin{eqnarray} \label{eq:depletion}
    \frac{dx_{ij}^a}{dt} &=& 
    \frac{1 - x_{ij}^a(t)}{\tau_r} - u_{ij}^a(t) \, x_{ij}^a(t) \, \delta(t - t^{i}_{sp}) \\    
    \frac{dx^{a}_{\text{astro}}}{dt} &=& 
    \frac{1 - x^{a}_{\text{astro}}(t)}{\tau_{r,\text{astro}}} - U_{\text{astro}} \, x^{a}_{\text{astro}}(t) \, \Theta([\Ca^{2+}]_{a} - \Cath) \:,
\end{eqnarray}
where $\tau_r$ and $\tau^{r, \text{astro}}$ are the recovery time constants for synaptic and astrocyte processes, respectively. 
In this model, we assume these values to be uniform across all processes.
The synaptic resource variable $x_{ij}^a(t)$ decreases instantaneously with every pre-synaptic spike $t^i_{sp}$, and recovers exponentially between spikes with time constant $\tau_r$. 
Similarly, the astrocyte resource variable $x^{a}_{\text{astro}}(t)$ depletes proportionally to gliotransmitter release, which occurs when astrocyte calcium concentration exceeds the threshold $\Cath$, and recovers with time constant $\tau_{r,\text{astro}}$.
This formulation ensures that both synaptic and astrocyte release dynamics are constrained by a limited pool of resources.

\subsection{Dimension two dynamics: the astrocytes}
 
Astrocytes possess two main compartments: the soma and the astrocyte processes, i.e. the terminal structures (leaflets) that directly interact with neuronal synapses and form the tripartite synapse \cite{Semyanov2021}. 
Each process can host localized calcium domains with faster dynamics compared to the soma \cite{Bazargani2016}. 
Calcium can diffuse from one process to another and to the soma where the signals from different processes are integrated, enabling a global calcium response {and} territorial synaptic modulation \cite{Araque2014}. 
This means that an astrocyte can modulate one synapse based on activity at other synapses within its territory of influence. 
Here we consider a simplified scenario where calcium diffuses instantaneously to the soma of the astrocyte, given {that} this is a much faster process, and we can thus focus on the integration effect of the astrocyte. 
 
We describe the state of astrocyte $a$ by its calcium concentration $[\Ca^{2+}]^a(t)$. 
Similarly to neuronal dynamics, when $[\Ca^{2+}]^a(t)$ crosses a threshold $\Cath$ we assume the astrocyte to become active and starts to release  gliotransmitters into its neighbouring synapses, modulating the presynaptic release probability $u_{ij}^a(t)$, as shown in the previous section (eqs. \ref{Eq_preact} and \ref{Eq_U_fin}). 
Unlike neuronal dynamics and previous works that model this as a discrete event \cite{Lenk2020, Pitta2022} (i.e. a spike), we treat gliotransmitter release as a continuous process. 
This is supported by experimental and computational evidence indicating that astrocyte calcium signals are slow (lasting from hundreds of milliseconds to several seconds or more) compared to the millisecond-scale action potentials in neurons \cite{Pasti1995, Verkhratsky2012, Sasaki2014, Manninen2020}. 
Therefore, we consider an astrocyte to be active as long as $[\Ca^{2+}]^a(t)$ remains above the threshold $\Cath$ (eqs. \ref{Eq_preact} and \ref{eq:depletion}). 
  
We consider in our model three main mechanisms driving $[\Ca^{2+}]^a(t)$ dynamics.
Firstly, $[\Ca^{2+}]^a(t)$ increases following each pre-synaptic spike in the synapses $ij$ modulated by astrocyte $a$. 
This occurs through the release of glutamate caused by the firing of the pre-synaptic neuron, which in turn leads to the synthesis of $IP_3$ (Inositol Trisphosphate, a signaling molecule) inside the astrocyte \cite{ip3metabo}. 
Following the formulation proposed by Ref. \cite{Lenk2020}, which simplifies earlier work \cite{DePitta_2008}, we assume that the concentration of $IP_3$ at the astrocyte processes of astrocyte $a$ that interact which synapse $ij$, $[IP_3]_{ij}^{a}$, increases instantaneously whenever the presynaptic neuron $i$ fires an action potential, namely 
\begin{equation}\label{eq:ip3_ija}
\frac{d[IP_3]_{ij}^a}{dt} = -\frac{[IP_3]_{ij}^a}{\tau_{IP_3}} + W_{ij}^a(t)\left[1 - [IP_3]_{ij}^a\right] \delta(t - t^{i}_{sp}) \: ,
\end{equation}
where the term 
\begin{equation}\label{eq:1-alpha}
W_{ij}^a(t) = (1 - \alpha) x_{ij}^a(t) u_{ij}^a(t)
\end{equation}
represents the coupling efficacy between the synapse $ij$ and the astrocyte process. 
Here, $(1 - \alpha)$ denotes the fraction of neurotransmitters that escapes the synaptic cleft and binds to receptors on astrocyte processes (such as glutamate metabotropic receptors).   
We note that this formulation constrains $[IP_3]$ between $0$ and $1$, serving both as a normalization and a simple way to incorporate saturation.  
As discussed above, we assume that $IP_3$ diffuses rapidly within the astrocyte, and thus we can write the total $IP_3$ concentration in astrocyte $a$, $[IP_3]^a$, as 
\begin{equation}
[IP_3]^a = \sum_{ij \in nn(a)} [IP_3]_{ij}^a(t) \:,
\end{equation}
where $nn(a)$ denotes the set of synaptic neighbors, i.e., all synapses adjacent to the astrocyte.

The second mechanism in the astrocyte calcium dynamics is calcium release or clearance away from the soma and astrocyte-synapse processes. 
Different mechanisms are at play in this processes, including e.g. SERCA pumps that transport calcium into the astrocyte's endoplasmatic reticulum (where it is not readily available for gliotransmission) \cite{Manninen2020}. 
Here, we simplify these processes to an effective exponential decay of $[\Ca^{2+}]^a$ over time, following approaches used in previous models for both synaptic and astrocyte calcium dynamics \cite{Graupner2012, Chindemi2022, Pitta2022}. 
Finally, the third mechanism of astrocyte calcium dynamics accounts for direct interaction between astrocytes. 
Astrocytes are interconnected through gap junctions, forming undirected astrocyte networks that enable calcium diffusion between cells \cite{Halassa2010}. 
Considering the three mechanisms together, we arrive at the following dynamics for the calcium concentration at astrocyte $a$:
\begin{eqnarray}
\frac{d[\Ca^{2+}]^a}{dt} &=& 
-\frac{[\Ca^{2+}]^a}{\tau_{\Ca}} 
+ \beta [IP_3]^a(t) 
+ \sum_{k \in ann(a)} D_{\Ca} \left( [\Ca^{2+}]^k - [\Ca^{2+}]^a \right), 
\end{eqnarray}\label{eq:calcium}
where $\beta$ has units of $ms^{-1}$, $\tau_{\Ca}$ is a decay constant that captures effective calcium clearance due to calcium pumps, and $D_{\Ca}$ is the diffusion coefficient modeling gap-junction-mediated calcium exchange with astrocyte neighbors $ann(a)$.

\begin{table}[!ht]
  \begin{center}
    %\label{tab:Parameters}
    \begin{tabular}{l|c|c}
      \textbf{Parameter} & \textbf{Symbol} & \textbf{Value} \\
      \hline
      Membrane time constant & $\tau_V$ & $20$ ms \\
      Membrane resistance & $R$ & $200$ M$\Omega$ \\
      Absolute refractory period & $t_a$ & $4$ ms \\
      Spike threshold potential & $V_{th}$ & $5$ mV \\
      Reset potential & $V_{reset}$ & $-5$ mV \\
      Synaptic deactivation time constant & $\tau_{in}$ & $4$ ms \\
      Absolute synaptic efficacy & $\mathcal{A}_{SE}$ & $3$ nA \\
      Fraction of neurotransmitter in the cleft & $\alpha$ & $0.25$–$0.975$ \\
      Neurotransmitter recovery time constant & $\tau_{r}$ & $100$ ms \\
      Gliotransmitter recovery time constant & $\tau^{as}_{r}$ & $100$ ms \\
      Steady-state gliotransmitter release probability & $U^{as}_{SE}$ & $0.1$ \\
      Steady-state neurotransmitter release probability & $U_{SE}$ & $0.1$ \\
      Astrocyte facilitation modulation parameter & $\varepsilon$ & $0.01$–$0.2$ \\
      IP$_3$ decay time constant & $\tau_{IP_3}$ & $6$ ms \\
      Calcium decay time constant & $\tau_{\Ca}$ & $100$ ms \\
      Astrocyte calcium diffusion coefficient & $D_{\Ca}$ & $0-10^{-1}$ $\text{ms}^{-1}$ \\
      Calcium increase rate via IP$_3$ & $\beta$ & $0.05-0.15$ $\text{ms}^{-1}$\\
      Synaptic facilitation time constant (pre-synaptic) & $\tau_{f}^{pre}$ & $200$ ms \\
      Synaptic facilitation time constant (astrocyte) & $\tau_{f}^{as}$ & $5000$ ms \\
      Gliorelease calcium threshold & $\Cath$ & $0.02-0.04$ $\mu$M\\
      External synaptic current amplitude & $A_{ext}$ & $0.3$ nA
    \end{tabular}
  \end{center}
   \caption{
   Description, symbols, and values of model parameters. To ensure physiologically plausible synaptic currents, $\mathcal{A}_{SE}$ is tuned such that the peak current from an individual synapse is of order $\mathcal{O}(10^{-1})$ nA. 
    For example, typical AMPA receptor-mediated synaptic currents are around $0.3$ nA \cite{Gerstner2014}. 
    Given that both $x_{ij}$ and $u_{ij}$ are on the order of $\mathcal{O}(10^{-1})$, it follows that $\mathcal{A}_{SE}$ should be of order $\mathcal{O}(1)$ nA to yield realistic current magnitudes.
    \label{tab:Parameters}
    }
\end{table}

\section{Results}
 To study in detail the effects of the tripartite synapse on synaptic and neuronal dynamics, we consider a minimal recurrent circuit of three excitatory neurons arranged in a ring, such that each neuron projects an excitatory (directed) synapse onto the following one, as shown in Figure \ref{fig:Exp_scheme} ({for simplicity, in the main part of this study we focus on this small recurrent network and in the final section we will explore the effect of increasing the circuit size}).
The network receives external input via {a spike train} arriving at one of the neurons, {namely the \emph{read-in neuron}}, and activity is read from the last neuron in the ring, {hereby refereed to as \emph{read-out neuron}}. 
As main control parameters of the model we consider the fraction $\alpha$ of neurotransmitters that remain available for the synapses -- such that $(1-\alpha)$ is the fraction of neurotransmitters recruited by the astrocyte -- and the frequency of incoming pulses at the read-in neuron.  

This neuronal circuit already exhibits various behaviors in the absence of astrocyte interaction, driven by local short-term {plasticity} (see Supp. Section 2, Figure A), including synfire activity, bistable (switching on-off) activity controlled by external stimuli, and bursting activity. 
As shown in Figure \ref{fig:Run_away_control}a, in this case the circuit is highly susceptible to entering a state of self-sustained activity (SSA) where the network is unresponsive to external stimuli -- equivalent to a supercritical state \cite{Obyrne2022}.
This occurs as soon as synaptic currents are large enough so that each pre-synaptic spikes triggers a post-synaptic one. 
Consequently, the firing rate of the read-in neuron remains high for all values of $\alpha$, and the SSA regime spans a large area of the parameter space (indicated by the black line in Figure \ref{fig:Run_away_control}a).  
As we go on to show (see also Supp. Section 2, Figure B), astrocyte modulation disrupts the SSA regime and extends the region where the system is responsive to external stimuli. 

{
In order to study the role of astrocyte modulation on preventing runaway activity, we focus here mainly on the case $\varepsilon<U_{SE}$. 
That implies, according to eq. \ref{Eq_U_fin}, that astrocyte regulation in the tripartite plasticity leads only to a decrease in neurotransmitter release probability, i.e., a sort of anti-facilitation. 
In a later section, we also consider the effect of positive astrocyte facilitation ($\varepsilon>U_{SE}$). 
To characterize the system dynamics, we measure the activity of the read-in (neuron $1$ in figure \ref{fig:Exp_scheme}) and read-out (neuron $3$ in figure \ref{fig:Exp_scheme}) neurons (respectively $\langle \nu^{in} \rangle$ and $\langle \nu ^{out} \rangle$) to characterize the possible functional benefits and side-effects of this negative astrocyte modulation. 
We compare different modulation schemes, in particular low and higher-order. In low-order schemes, each astrocyte modulates only one synapse. In higher-order schemes, on the contrary, each astrocyte modulates several synapses (but still each synapse can only be modulated by one astrocyte). 
This corresponds to the actual case in neuronal-astrocyte networks, and naturally induces higher-order interactions. }
Finally, we will consider the possibility of astrocyte-astrocyte interactions through gap-junctions (eq. \ref{eq:calcium}) and who this affects the dynamics of the system, and whether these interactions can recover the effect of actual higher-order astrocyte regulation in low-order schemes. 

\subsection{Higher-order modulation prevents runaway excitatory activity.}

\begin{figure}[!ht]
    \centering
    \includegraphics[width=0.85\linewidth]{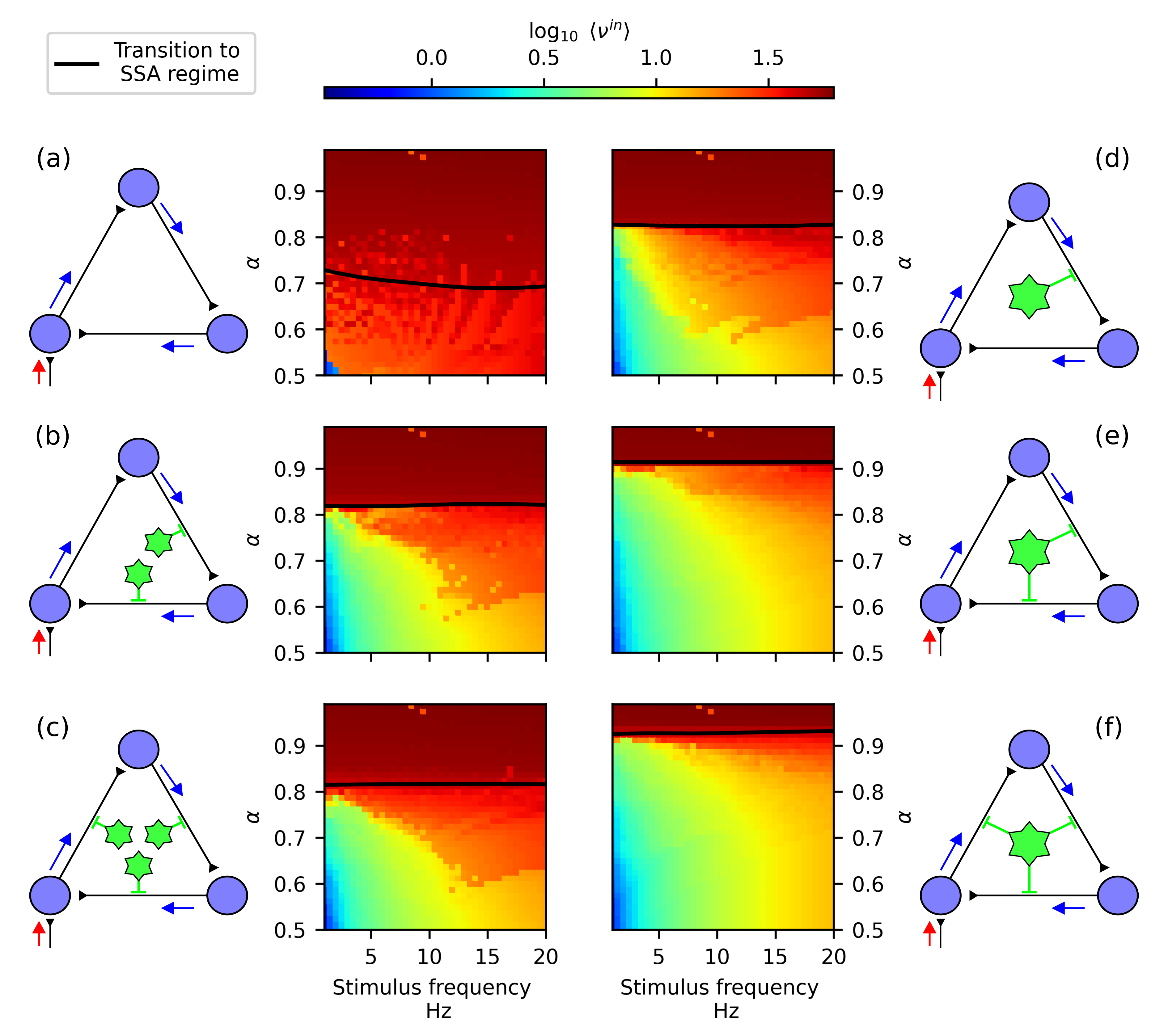}
    \caption{\textbf{Astrocyte higher-order modulation extends the responsive region of the neuronal circuit.} 
     We show network activity through the average firing frequency of the read-in neuron $\left\langle \nu^{in} \right\rangle$ (heat-map diagrams) as a function of the fraction of neurotransmitters that remain in the synaptic cleft ($\alpha$), and of the stimulus frequency. 
    Each panel corresponds to a different astrocyte configuration as indicated by the corresponding graph diagram: 
    (a) no astrocytes; 
    (b–d) respectively one, two, and three astrocytes {that} interact with single synapses (low-order regulation); 
    (e–f) one astrocyte {that} interacts with multiple synapses (higher-order regulation).
    In the graph diagrams blue circles stand for neurons, black lines for synapses (with the round end pointing towards the post-synaptic neuron), and green stars for astrocytes.
    Astrocyte-synapse interactions are indicated by green short lines.
    The read-in neuron that receives the external stimulus and which activity is shown in the heat-maps is indicated by a red arrow.     
    In each diagram, the black line indicates the transition to the SSA regime. This line marks the critical value of $\alpha$ above which the synaptic currents emitted by neurons are always sufficient to trigger spikes in their postsynaptic partners, thereby sustaining activity within the circuit. For a detailed description of how the transition line is computed, see Supp. Section 4.
    Network parameters were set to the default values listed in Table \ref{tab:Parameters}, and we set $\varepsilon=0.01$, $\Cath=0.04$, $\beta=0.05$. 
    }
    \label{fig:Run_away_control}
\end{figure}

We assess the regulatory effect of astrocyte modulation on activity propagation under different schemes (see Figure \ref{fig:Run_away_control}), including \emph{low-order} and \emph{higher-order} ones. 
%In low-order schemes (panels b-d) each astrocyte modulates only one synapse, resulting in localized modulation. 
%On the contrary, in higher-order modulation schemes (panels e-f) an astrocyte modulates several synapses concurrently, thereby implementing an intrinsically higher-order link-to-link interaction in the neuronal circuit. % Quito esto porque ya lo decimos justo arriba
Our results show that the presence of even a single tripartite synapse is sufficient to regulate activity-propagation and significantly extend the area of the \emph{responsive} phase in the parameter space, in particular in the control parameter $\alpha$ (panels b-f).
Notably, this effect is greatest when astrocytes act in a higher-order scheme, that is, when a single astrocyte modulates multiple synapses (panels e,f), rather than astrocytes acting locally at the single-synapse level (panels b,c). 
In the higher-order scheme, astrocytes integrate the signals of different synapses. 
This not only leads to a faster $[\Ca^{2+}]$ increase in the astrocyte and quicker gliotransmitter release, but this release also occurs simultaneously across all adjacent synapses, thus allowing for a more effective modulation of neuronal activity. 
These results highlight the role of higher-order interactions, as mediated by astrocytes in the tripartite synapse, in regulating excessive activity at the node and edge level, thus expanding the responsiveness of the system of external stimuli. 

\begin{figure}[!ht]
    \centering
    \includegraphics[width=\linewidth]{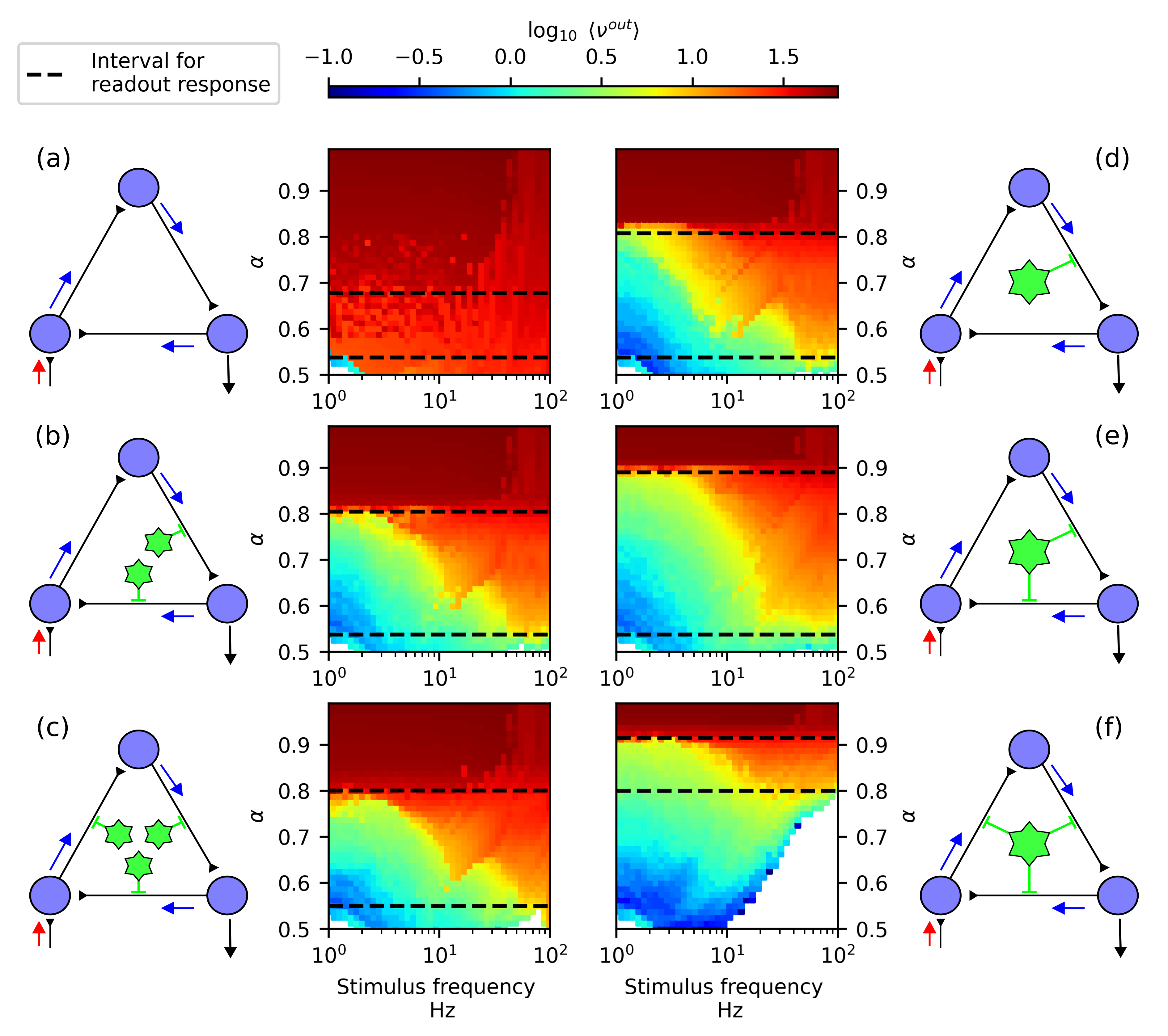}
\caption{\textbf{astrocyte interaction scheme determines the adequate read-out neuron response interval.}  
We show network activity through the average firing frequency of the read-out neuron {(indicated by a black arrow in the graph diagrams) $\left\langle \nu^{out} \right\rangle$.} 
All other details are as in Figure \ref{fig:Run_away_control}.
The region between black dashed lines indicates the range of $\alpha$ values where the read-out neuron exhibits an adequate response. 
Parameters: $\varepsilon = 0.01$, $\Cath = 0.04$, $\beta = 0.05$. Other simulation parameters are listed in Table \ref{tab:Parameters}.}
    \label{fig:Outread}
\end{figure}

\subsection{Higher-order tripartite plasticity improves read-out neuron response to stimuli.}

To explore possible side effects of astrocyte regulation, we consider in Figure \ref{fig:Outread}  the average firing-rate of the read-out neuron, {namely $\langle \nu^{out} \rangle$,} in the same modulation schemes as considered in the previous section (Figure \ref{fig:Run_away_control}).
In the absence of astrocyte regulation (panel a), {$\langle \nu^{out} \rangle$} is high except for a small region for small $\alpha$ and small stimulus frequency (below $2$Hz), for which the read-out neuron does not fire.  
In this regime a spike of the pre-synaptic neuron is not enough to cause a spike of the post-synaptic neuron, and STP does not effectively operate at such low frequencies due to the typical timescales of plasticity mechanisms \cite{Protachevicz2024}.
Otherwise the average firing rate is high as for the pre-synaptic neuron in Figure \ref{fig:Run_away_control}a, indicating low sensitivity of the circuit to the stimulation rate and its susceptibility to falling into the SSA regime.
Similarly to the read-in neuron case, the presence of at least one tripartite synapse is sufficient to modulate activity, significantly increasing the read-out neuron's sensitivity to external stimuli.
To quantify the ability of the circuit to respond to external stimuli, we defined the adequate response interval as the range in $\alpha$ for which the read-out neuron responds proportionally to stimulus frequencies ($1-100$Hz), and for which the circuit is not in the SSA regime. A comparison between circuit activity in the safe regime with ``adequate" response vs SSA regime is presented in Supp. Section 3. 
This is shown in Figure \ref{fig:Outread} as the area between the two black dashed lines. 
As shown in the figure, the addition of astrocyte modulation, either in a low- or higher-order setting, expands the adequate response interval except for the case where a single astrocyte regulates all three synapses (panel f), which we discuss in detail in the next paragraph. 
The configuration with the largest adequate-response interval (panel e) corresponds to higher-order regulation by a single astrocyte of the two internal synapses (synapses $1\to2$ and $2\to 3$).

In the higher-order case in which the astrocyte regulates all three synapses (panel f), we observe that the read-out neuron is silent for high stimulation frequencies, and the adequate-response interval moves to higher $\alpha$ values and shrinks in extension. 
These effects are only very weakly present in the corresponding low-order case in which there are three astrocytes, one regulating each synapse (panel c). 
Notably, as we mentioned above this phenomenon is not observed in the higher-order case in which the astrocyte does not regulate the first synapse (panel e). 
This is because the activation of the first synapse ($1\to 2$) can be independent to the internal dynamics of the circuit, as it is driven by the external stimulus which causes neuron $1$ to fire. 
The repeated activation of synapse $1\to 2$ maintains calcium concentration at astrocyte high (through the associated $IP_3$ currents, see eq. \ref{eq:ip3_ija}) and the astrocyte remains active for longer time, thus reducing synaptic response (by reducing the release probability, eq. \ref{Eq_U_fin}) and disrupting activity propagation before it arrives to the read-out neuron (for more details of this phenomenon in different modulation schemes see Supp. Section 5). 
A similar but significantly weaker effect occurs in the case of low-order modulation (panel c), but it is noticeable only for very small $\alpha < 0.55$ and very high stimulus frequencies ($>50$ Hz).

\begin{figure}[!ht]
    \centering
    \includegraphics[width=\linewidth]{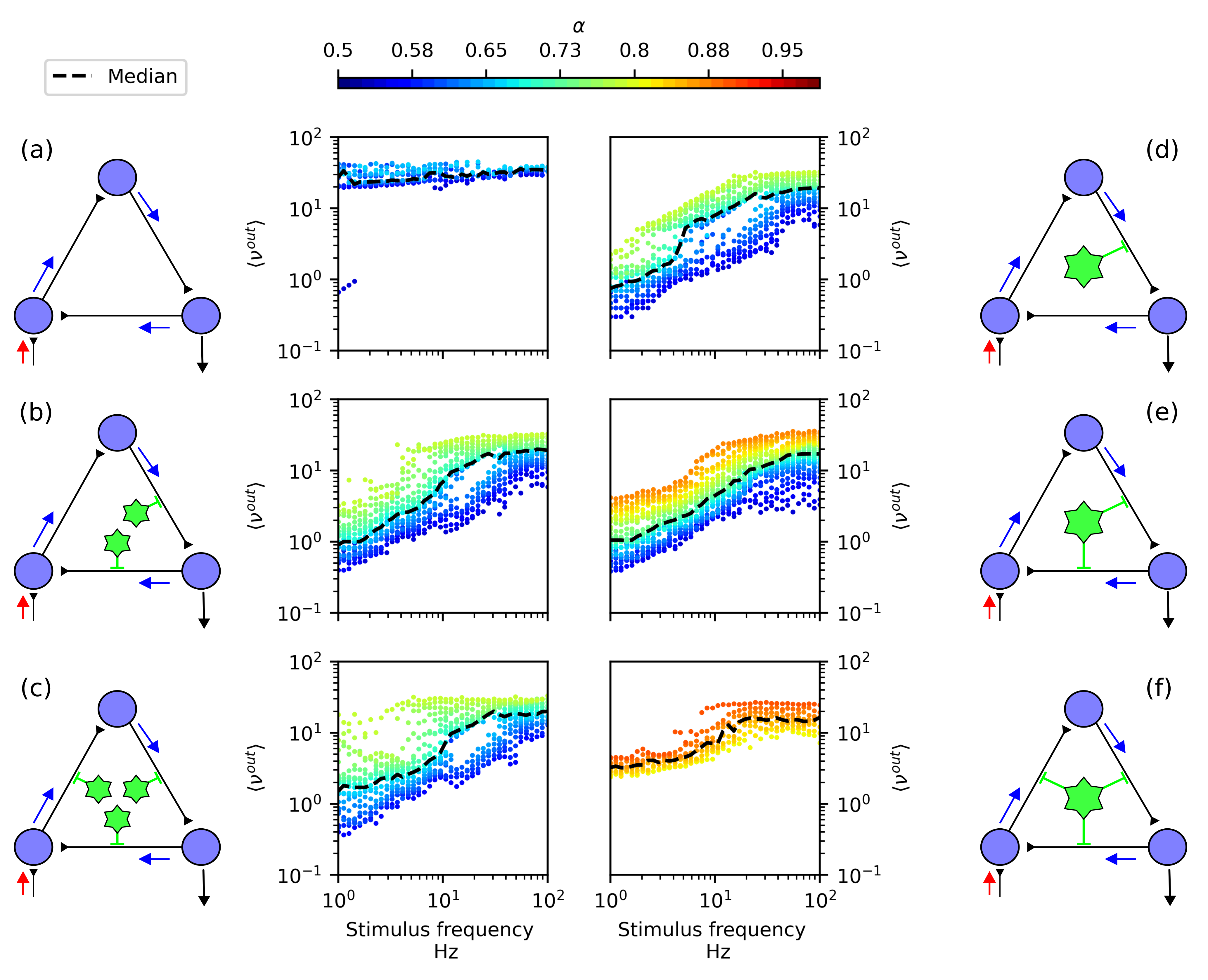}
    \caption{\textbf{Neuronal circuit response improves under higher-order astrocyte modulation of internal synapses}. 
    We show the same data as in Figure \ref{fig:Outread} restricted to the adequate-response regime, for each interaction scheme. 
    Here we show {$\langle \nu^{out}\rangle$} as a function of the {stimulus frequency}. 
    Each data-point corresponds to a value of $\alpha$, as indicated by the color-code (see color-bar). 
    The median firing rate for each stimulus frequency is shown by the black dashed line. 
    Parameter values were $\varepsilon=0.01$, $\Cath=0.04$, $\beta=0.05$ and $D_{\Ca}=0$.
    Other simulation parameter values are listed in Table \ref{tab:Parameters}.} 
    \label{fig:TransferFunction}
\end{figure}

In order to study the adequate response regime in more detail, in Figure \ref{fig:TransferFunction} we show $\langle \nu^{out}\rangle$ again, but now only for $\alpha$ values in the adequate response interval identified in Figure \ref{fig:Outread}, and as a function of the stimulus frequency.
The presence of astrocyte modulation increases the range of values of $\langle \nu^{out}\rangle$ (panels b-f). 
Low-order schemes lead to a bimodal distribution of response firing rates for input frequencies between 1-30 Hz, indicating that small changes in $\alpha$ could produce an abrupt increase in firing rate, making the circuit less stable with respect to external stimulation (panels b-d). 
The higher-order case in which only internal synapses ($2\to 3$ and $3\to 1$) are modulated (panel e) shows the largest frequency-range of $\langle \nu^{out}\rangle$. 
Moreover, it is the only case displaying a smooth response both in $\alpha$ and as a function of the input frequency, indicating a more robust and predictable response of the neuronal circuit. 
Conversely, the higher-order case where the three synapses are regulated (panel f) shows a reduced frequency-range of  $\langle \nu^{out}\rangle$ (which also expands a smaller range in $\alpha$ as discussed above). 

These results highlight a simple yet important point: for tripartite plasticity to be beneficial in a recurrent network such as the one studied here, modulation should only {apply to} \emph{internal} synapses, i.e. synapses that are activated by the network’s own activity. 
When astrocytes interact with synapses not regulated by the internal activity of the circuit, it means, synapses that will be activated by external factors, then excessive modulation can occur, undermining the benefits of astrocyte involvement. 
As we discuss in following sections, increasing $\varepsilon$  -- though still below $U_{SE}$ -- mitigates the effects of excessive modulation (Figure \ref{fig:Modulation_Outread}). 
However, achieving optimal modulation would require fine-tuning of $\varepsilon$, which is undesirable from any standpoints.

\subsection{On whether higher-order modulation differs from low-order modulation.}

\begin{figure}[!ht]
    \centering
    \includegraphics[width=\linewidth]{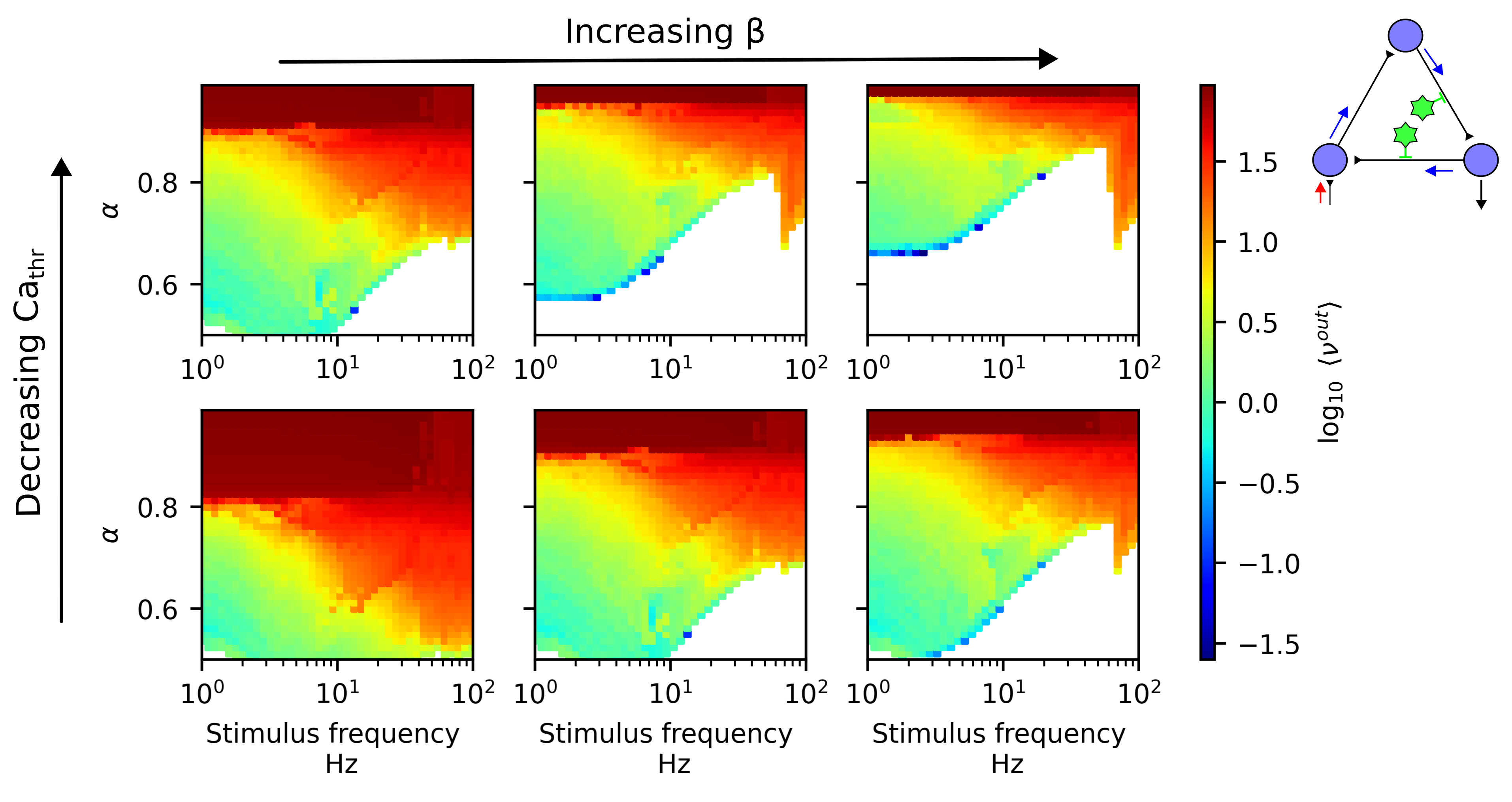}
    \caption{
    \textbf{Low-order modulation cannot replicate higher-order effect}. 
    We show the firing frequency of the read-out neuron $\left\langle \nu^{out}\right\rangle$ under a low-order interaction scheme as shown by the graph top-right inset (two uncoupled astrocytes regulating {the internal synapses} synapses $2\to 3$ and $3 \to 1$ respectively) {for different values} of the parameters $\beta$ and $\Cath$ controlling the astrocytes' calcium dynamics (eq. \ref{eq:calcium}).
    In particular, $\beta = \{0.05, 0.1, 0.15\}$ for left to right, and $\Cath = \{0.04, 0.02\}$ (bottom to top). 
    The reference values (considered in the previous figures) are $\beta=0.05$ and $\Cath=0.04$.    
    Gliotransmission is facilitated both by increasing $\beta$ and/or decreasing $\Cath$. 
    Here $\varepsilon=0.01$ and all other parameters are as listed in Table \ref{tab:Parameters}.
    }
    \label{fig:Gliorelease_high}
\end{figure}

In our set-up, low- and higher-order regulation schemes differ in the number of synaptic inputs received by each astrocyte, which {always equals} one in the low-order schemes, and two or three in the higher-order ones. 
This leads to lower $IP_3$ production, lower calcium concentration, and thus reduced astrocyte modulation {in the low-order schemes}. 
This raises the question of whether the differences observed in Figures \ref{fig:Run_away_control} and \ref{fig:Outread} between low-order (panels b and c) and higher-order (panels e and f) regulatory schemes are simply due to a different scaling of the calcium levels, rather than a true effect of higher-order regulation.
To test this, in Figure \ref{fig:Gliorelease_high} we performed additional analyses for the low-order modulation scheme shown in panel (c) of Figures {\ref{fig:Run_away_control} and \ref{fig:Outread}}, that is, the case where two different astrocytes regulate {the internal} synapses $2\to 3$ and $3\to 1$, respectively. 
In particular, {we analyzed how the parameters $\beta$ and $\Cath$ controlling the astrocytes' calcium dynamics affect the $\langle \nu^{out}\rangle$ diagrams.} 
In particular, $\beta$ controls the level of intracellular $[\Ca^{2+}]$ elevation due to $IP_3$ production, while $\Cath$ sets the calcium threshold required for gliotransmitter release (eq. \ref{eq:calcium}). 
Gliotransmission is facilitated both by increasing $\beta$ and/or decreasing $\Cath$. 
In Figure \ref{fig:Gliorelease_high} we systematically increase $\beta$ and decrease $\Cath$, starting from their reference values, thereby enhancing gliotransmission. 
We find that in either case the onset of the SSA regime (dark-red region) shifts towards higher $\alpha$ values, partially mimicking the behavior observed in the higher-order scheme of Figure \ref{fig:Outread}(e). 
However, in neither case do we recover the adequate modulatory regime found under the higher-order regulation scheme. 
On the contrary, by increasing gliotransmission in the low-order case we find that it disrupts synaptic signal propagation prematurely, and the readout neuron becomes unresponsive to external stimuli (white region) already for smaller frequency inputs. 
Thus, the modulatory benefits seen in the higher-order setup cannot be replicated by simply amplifying astrocyte calcium currents {in low-order schemes}. 
This underscores the functional significance of astrocyte-mediated higher-order interactions in modulating synaptic activity at the circuit level.

\subsection{Positive modulation by astrocytes.}

\begin{figure}[!ht]
    \centering
    \includegraphics[width=\linewidth]{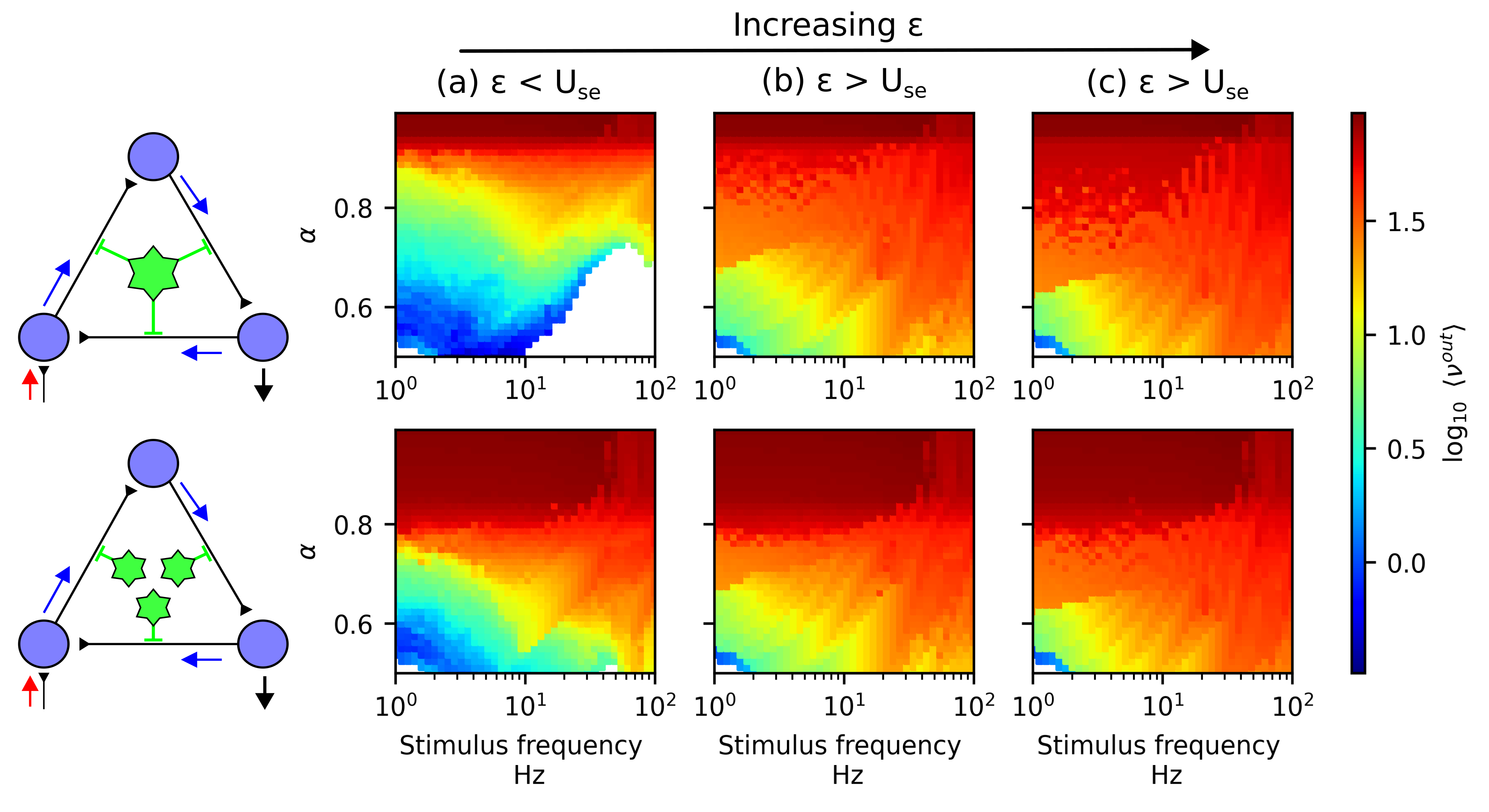}
    \caption{\textbf{From negative to positive astrocyte modulation (effect of $\varepsilon$).}
    Effect of the modulation type (controlled by $\varepsilon$) for a {higher}-order (top panels) and the corresponding low-order (bottom panels) schemes, as indicated by the graph diagrams. 
    {Panel (a) corresponds to decreased facilitation, $\varepsilon = 0.05 < U_{SE}$, whereas panels (b) and (c) correspond to increased facilitation ($\varepsilon > U_{SE}$), respectively with $\varepsilon = 0.15$ and $0.2$.}
    The reference value (used for the previous figures) is $\varepsilon = 0.01$.
    Parameter values were $\Cath=0.04$, $\beta=0.05$ and $D_{\Ca}=0$. Additional simulation parameter values are provided in Table \ref{tab:Parameters}.
    }
    \label{fig:Modulation_Outread}
\end{figure}

In the previous analyses we have assumed that gliotransmission has a depressive modulatory effect on synaptic transmission. 
More generally, however, gliotransmission can also have a facilitating effect. 
This is captured in the model by the parameter $\varepsilon$: for 
$\varepsilon < U_{SE}$ gliotransmission decrease facilitation, whereas for $\varepsilon > U_{SE}$ it increase facilitating. 
For $\varepsilon = U_{SE}$, gliotransmission has no effect, a case know as occlusion \cite{DePitta2019}. 
In Figure \ref{fig:Modulation_Outread} we compare decreased and increased facilitation both in the lower-order (top panels) and higher-order (bottom panels) interaction schemes. 
We find that, firstly, the extension of the SSA regime in $\alpha$ (dark-red area) is not drastically affected by type of astrocyte modulation (i.e. decreased or increased facilitation), but only by the presence of astrocytes and the interaction scheme. 
That is, the extension of the SSA regime in parameter space shrinks under the presence of higher-order astrocyte regulation (top panels), regardless of whether gliotransmission decrease or increase facilitation.
Thus, the mere competition for activate pre-synaptic neuroreceptors (quantified by $\gamma_{as}$ and $\gamma_{pre}$) is enough to prevent (or significantly reduce) runaway activity in the neuronal circuit.
Secondly, we find that facilitating astrocyte modulation does lead to higher firing frequencies in the read-out neuron, as one might expect, and thus the region of excess suppression leading to unresponsiveness of the read-out neuron (white regions in the right-had side of the diagrams) is significantly reduced. 
This effect already appears by increasing $\varepsilon$ even for $\varepsilon<U_{SE}$, as it can be noticed by comparing panels (a) of Figure \ref{fig:Modulation_Outread} (with $\varepsilon=0.05<U_{SE}$) to panels (c) and (f) of Figure \ref{fig:Outread} (with $\varepsilon=0.01<U_{SE}$ and all other details equal).
This suggests that fine-tuning $\varepsilon$ could enhance the benefits of astrocyte modulation, although such precise tuning may not be plausible in biological systems.

\subsection{Astrocyte diffusive network.}

\begin{figure}[!ht]
    \centering
    \includegraphics[width=\linewidth]{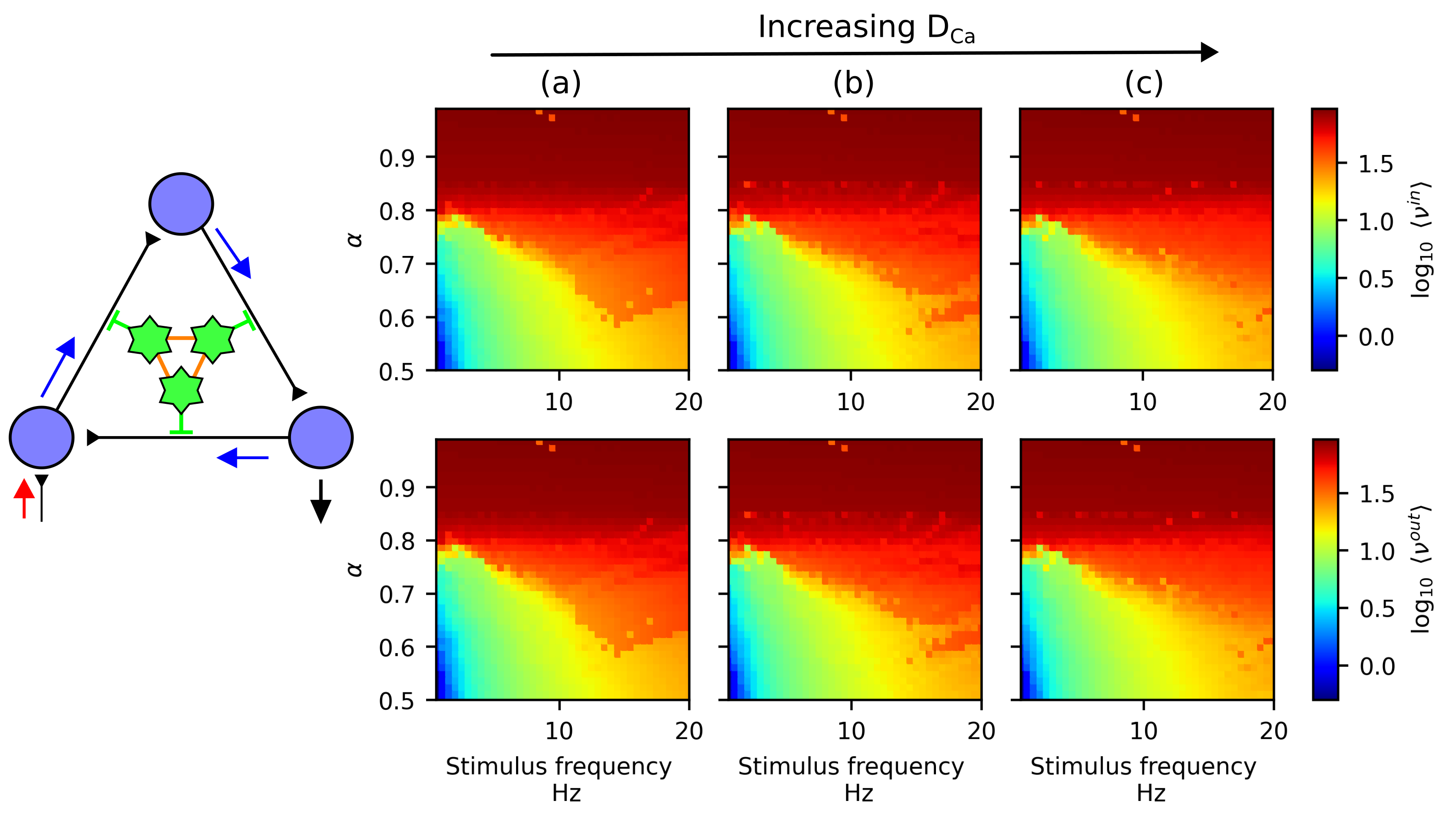}
    \caption{
    \textbf{Astrocyte diffusion-network does not recover higher-order modulation schemes}. 
    Here we consider the low-order modulation scheme with three astrocytes, such that each astrocyte modulates one synapse (corresponding to panel (c) in Figures \ref{fig:Run_away_control} and \ref{fig:Outread}), as shown by the graph diagram. 
    In this case the astrocytes are coupled via gap-junctions (eq. \ref{eq:calcium}) as represented by red lines in the graph cartoon. 
    Each column corresponds to a different diffusion coefficient $D_{\Ca}$, namely $D_\Ca=10^{-3} \text{ms}^{-1}$ (panels a), $10^{-2} \text{ms}^{-1}$ (panels b), and $10^{-1} \text{ms}^{-1}$ (panels c). 
    For each panel, we show via a colormap the average firing rate of (top row) the read-in neuron $\left\langle\nu^{in}\right\rangle$ and (bottom row) the read-out neuron $\left\langle\nu^{out}\right\rangle$. 
    Parameter values were $\varepsilon=0.01$, $\Ca_{th}=0.04$ and $\beta=0.05$ in $\text{ms}^{-1}$. Other simulation parameter values are listed in Table \ref{tab:Parameters}.}
    \label{fig:With_difussion}
\end{figure}

The astrocyte model also allows us to introduce coupling between astrocytes through a diffusive term representing gap-junction interactions. 
Generally, this can have non-trivial effects on the emergent dynamics of the system, particularly in large systems. 
Here, and in line with the rest of the study, we consider in particular the question of whether astrocyte-coupling in low-order schemes, in which each astrocyte modulates only one synapse, can replicate the findings of higher-order schemes, in which an astrocyte modulates several synapses concurrently. 
In Figure \ref{fig:With_difussion}, we consider the low-order scheme of three astrocytes, each modulating a synapse, corresponding to panel (c) of Figures \ref{fig:Run_away_control} and \ref{fig:Outread}. 
In this case, we couple the three astrocytes via pair-wise gap-junctions {(according to eq. \ref{eq:calcium})}, as indicated in the figure, and consider increasing diffusion coefficients $D_{\Ca}$. 
Our findings show that calcium diffusion across astrocytes has a small effect that does not significantly alter the phase diagram {of the system}, and, in particular, is not able to recover the effects of true higher-order modulation. 
When a single astrocyte modulates several synapses, it receives and integrates inputs from all of them, resulting in enhanced gliotransmission. 
Therefore, it is an aggregative, rather than diffusive, effect, and it cannot be replicated via diffusive coupling.

\subsection{Astrocyte modulation in {larger networks}.}

\begin{figure}[ht!]
    \centering
    \includegraphics[width=\linewidth]{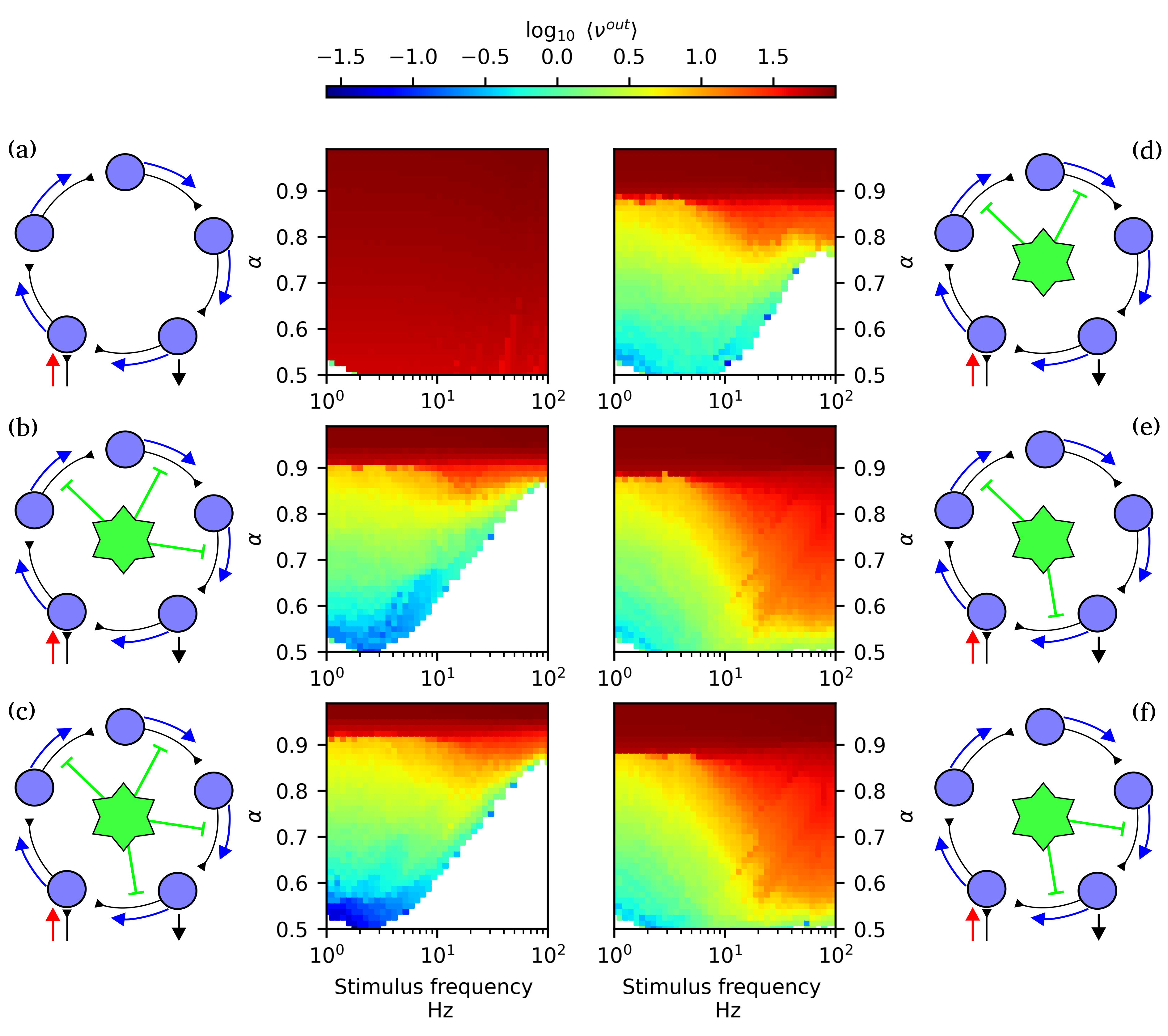}
    \caption{\textbf{Astrocyte modulation in a larger cycle of five neurons.} 
    Colormaps indicate the average firing rate of read-out neuron $\left\langle \nu^{out} \right\rangle$. 
    Each panel corresponds to a different interaction scheme, as indicated by the graph diagrams on the top row. 
    Namely, no astrocytes (panel a); and a single astrocyte modulating three internal synapses (panel c); all four internal synapses (panel c); or just two internal synapses (panels d,e,f), where in (panel d) astrocite does not modulate the recurrence synapse and in (panels e and f) it modulates the recurrence one from the read-out to the read-in neuron, and a second one. 
    Simulation with $\varepsilon=0.01$, $\Cath=0.04$, $\beta=0.05$ and $D_{\Ca}=0$. Other simulation parameter values are listed in Table \ref{tab:Parameters}.
    %q{In the other figures we put the labels column-wise (so the first column would be a, b, c, instead of a, c, e), can we do that here? Also, what is the rationale for the order of the panels? Since we start with 0 regulations I would then go up to 2, 3, 4, instead of going in decreasing order.}
    %q{I would prefer to have a centered pentagon, with the read-in and read-out neurons both at the bottom, like we did for the triangle, so that link is horizontal, and the third neuron is centered at the top. Then also the black arrow can be vertical like the red one. }
    }
    \label{fig:Five_neurons}
\end{figure}

In {the main part of this study we have considered a simple system made by a three-neuron recurrent circuit. 
To gain a glimpse on the effects of higher-order astrocyte modulation on larger systems, we consider here {larger networks in the form of $N$-node cycles under different modulation schemes. 
Firstly, in Figure \ref{fig:Five_neurons}} we consider $N=5$ and} four interaction schemes:
(a) no astrocyte regulation; 
(b) a single astrocyte modulating all internal synapses; and 
(c–d) a single astrocyte modulating two synapses, one of which is the recurrence synapse connecting the read-in and read-out neurons.
We find that, in the absence of astrocyte regulation (panel a), the system now enters the SSA regime for almost any value of the control parameters. 
This behaviour is {due to the interplay between the intrinsic time-scales of short-term plasticity recurrent pulses that travel through the cycle, which cause the system to enter a resonance state insensitive to the details of the external stimulus. 
For more details of this phenomenon see Supp. Section 6.}

A responsive operating regime becomes possible only in larger cycles through astrocyte modulation.
To analyze the effect of particular astrocyte interaction schemes, we focus on those not regulating the input synapse, as we discussed above.
We find that, when the astrocyte modulates all synapses but this one ($n=4$ interacting synapses, panel b),
%Notably, when the astrocyte modulates several synapses (panel b), 
excessive modulation arises as the frequency of the stimulus increases (white region in the right-hand side of the diagrams) that is akin to the behavior in panel (f) of Figure \ref{fig:Outread} and that ultimately disrupts the network’s responsiveness to external stimuli. 
In this case, excessive astrocyte ``inhibition" (through decreased facilitation) arises from its interaction with several synapses, which rapidly rises $[\Ca^{2+}]$ in the astrocyte.
Given that the astrocyte regulates all synapses concurrently, it stops 
%In short, interaction with several synapses produces an excess of $IP_3$ current which leads to over-modulation and kills 
activity propagation before it reaches the read-out neuron.

The case of intermediate higher-order modulation, i.e., when the astrocyte modulates two synapses {($n=2$), one of which is the recurrence one and the other is an internal synapse,} achieves optimal response of the read-out neuron {(panels e and f)}. 
In this manner, the astrocyte effectively integrates information about overall network activity through the later synapse, while directly modulating recurrent excitability with the former. This interplay enables an optimal operational regime that balances propagation and stability.
Notably, when the astrocyte only modulates internal synapses (panels c for $n=3$ and d for $n=2$), excessive modulation still occurs.
%The dynamics of the read-out neuron are nearly identical across both configurations. %In both of them, the astrocyte modulates the recurrence or final synapse (from the read-out to the read-in neuron) together with an additional internal synapse. 
%q{Q: In panel b we also regulate the recurrence synapse but we still get the unresponsive regime because there are too many regulated synapses ($n=4$). 
%In panel c ($n=3$) it is not clear whether we get the unresponsive regime because we don't modulate the recurrence synapse or because there are too many regulated synapses. 
%Why do we show this case? For me it would be more logical to keep the recurrence synapse, and we can focus on the effect of the number of regulated synapses. 
%Then for $n=2$ we focus  on the discussion of which synapses. 
%Going back to the effect of $n$, should we have included some normalization of the IP3 currents by the number of adjacent synapses?
%}

\begin{figure}
    \centering
    \includegraphics[width=0.9\linewidth]{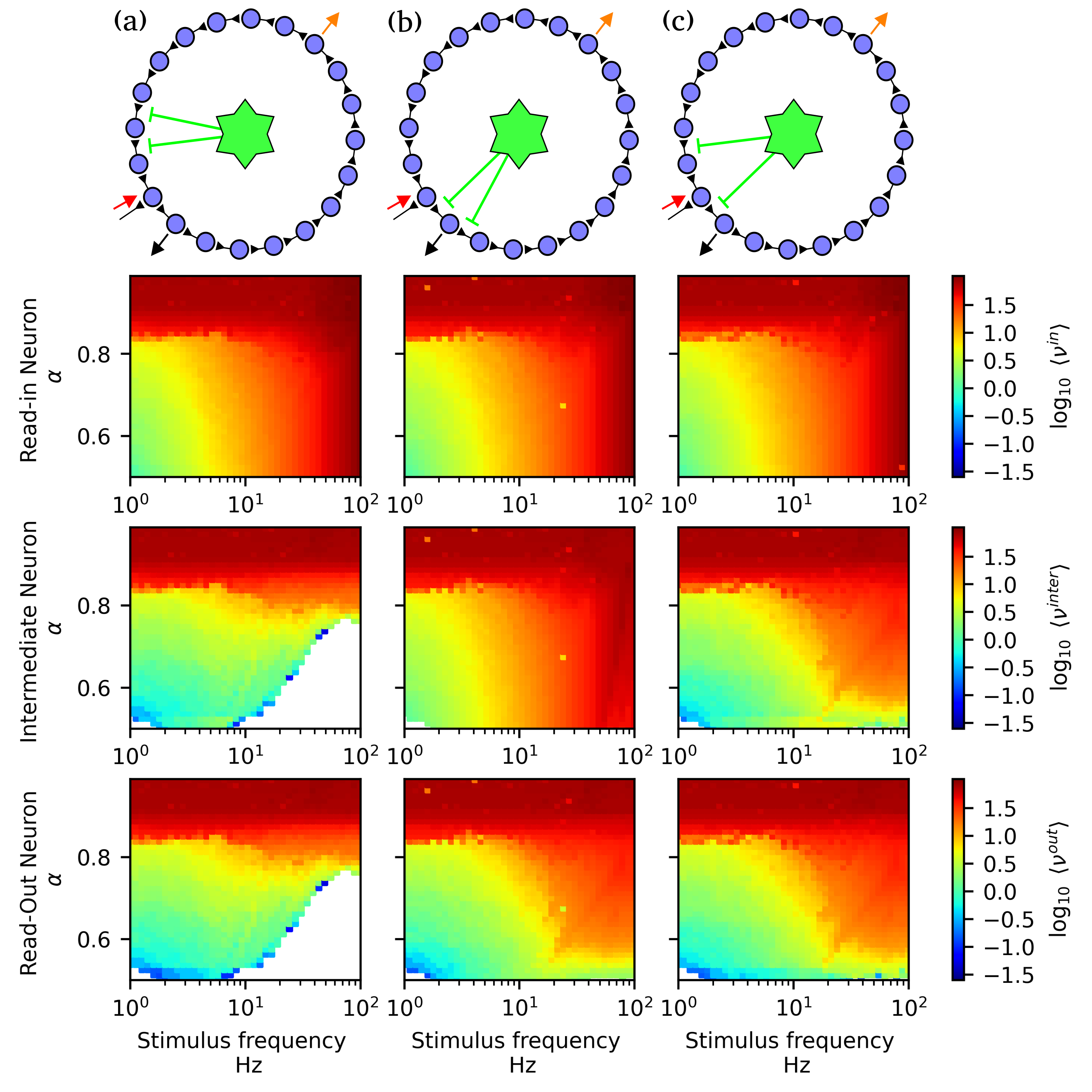}
    \caption{
    \textbf{{Interaction architecture shapes internal response in larger systems.}} %Internal neuron reproduces read-in or read-out response depending on the position of astrocyte-modulated synapses.}  
    Colormaps indicate the average firing rate of the read-in $\left\langle \nu^{\text{in}} \right\rangle$ (top), intermediate $\left\langle \nu^{\text{inter}} \right\rangle$ (middle), and read-out $\left\langle \nu^{\text{out}} \right\rangle$ (bottom) neurons. 
    Each column corresponds to a different astrocyte interaction scheme, illustrated by the graph diagrams at the top. Red, orange, and black arrows indicate the read-in, intermediate, and read-out neurons, respectively. 
    In all cases, the astrocyte modulates two synapses: 
    (a) {not including the recurrence synapse; 
    (b) the recurrence synapse and the previous synapse; and
    (c) the recurrence synapse and the first internal synapse}. 
    Simulations were performed with $\varepsilon = 0.01$, $\Cath = 0.04$, $\beta = 0.05$, and $D_{\Ca} = 0$. Other simulation parameters are listed in Table~\ref{tab:Parameters}.}
    \label{fig:Twenty_neurons}
\end{figure}

{To gain more insight into the effects of regulating the recurrence synapses, }
extended the analysis of astrocyte modulation of two synapses to a larger {cycle} consisting of $20$ excitatory neurons {in Figure \ref{fig:Twenty_neurons}}. 
{We considered in particular three schemes: two internal synapses (panel a), the recurrence synapse and the one before it (panel b), and the recurrence synapse and an internal synapse close to the read-in neuron (panel c).
We consider in this case}
%In Figure \ref{fig:Twenty_neurons}, we show 
the average firing rates of the read-in neuron {$\langle \nu^{in}\rangle$ (second row of panels)}, an intermediate neuron, {indicated by the orange arrow in the graph panels, $\langle \nu^{inter}\rangle$ (third row of panels)}, and the read-out neuron {$\langle \nu^{out}\rangle$ (forth row of panels)}.
As expected from {our} previous results, the read-in neuron exhibits a safe response to external stimuli of any frequency and across a broad range of $\alpha$ values, due to higher-order astrocyte modulation (see top row of Figure~\ref{fig:Twenty_neurons}).
As before, we observe that optimal responsiveness {of} the read-out neuron occurs when the recurrence synapse is modulated {by the astrocyte (panels b and c), regardless of which other synapse is modulated.
As expected, when the two regulated-synapses are internal and situated early on in the cycle, 
} %independent{ly of} the position of the co-modulated synapse.
In contrast, when the astrocyte modulates only two of the first synapses in the circuit (panel a),
signal propagation is suppressed for a large region of the parameter space and stimulus frequency, {and the read-out neuron is unresponsive (white region in the right-hand side of the read-out neuron diagram).} 

{To investigate the robustness of these results over the network, we consider now the activity of the internal synapse.}
Interestingly,  when the recurrence synapse is modulated {the activity of the internal synapses does depend on the location of the second modulated synapse.} %, the behavior of the intermediate neuron depends on which second synapse is co-modulated. 
{When}
the astrocyte modulates the last two synapses in the {cycle (panel b)}, activity in the early part of the circuit is weakly modulated, and the intermediate neuron {behaves similarly to the read-in neuron.}
%exhibits a response similar to that of the read-in neuron—higher firing rates across a wide frequency range (see Figure~\ref{fig:Twenty_neurons} middle row, column b). 
In contrast, if the {second synapse modulated by the astrocyte is located early-on in the cycle (panel c), the behavior of the intermediate neuron resembles now that of the read-out neuron (panel c).} %astrocyte modulates both the recurrence synapse and one of the first synapses in the circuit, the intermediate neuron instead mirrors the behavior of the read-out neuron—lower firing rates and stronger attenuation of stimulus response (see Figure~\ref{fig:Twenty_neurons} middle row, column c). 
This result suggests that the astrocyte’s spatial scope of integration -- whether it spans the beginning and end of the circuit, {i.e., accessing} non-local information about the network state, or only the end, {i.e., accessing only local information} -- {influences activity propagation through the cycle, and the responsiveness of the internal neurons to the external stimuli}.

\section{Discussion and conclusions} 

In recent years, there has been a growing recognition of the active role astrocytes play in shaping neuronal network dynamics \cite{Miguel-Quesada2023}. 
Far from being passive support cells, astrocytes are now understood to participate in a wide range of processes, from neural circuit development \cite{Clarke2013} to regulation of neural circuit activity \cite{Oliveira2022}. 
In particular, their involvement in synaptic plasticity -- through the release of gliotransmitters and the modulation of pre- and postsynaptic activity -- has attracted increasing attention \cite{Ota2013,Squadrani2024}. 
{For instance, glutamate clearance, primarily carried out by astrocytes \cite{Radulescu2022}, is essential for normal brain function and to prevent excitoxicity which has been hypothesized to play a role in neurodegenerative diseases including amyotrophic lateral sclerosis, Alzheimer's disease and Huntington's disease \cite{glutamateclearance2015}.
Furthermore, gliotransmission can activate extrasynaptic inhibitory receptors through astrocyte GABA release, a mechanism that, when dysregulated, has been linked to Alzheimer’s disease \cite{Jo2014}. 
These and other effects highlight astrocyte-neuron interactions as key regulators of both physiological and pathological behaviors in neuronal media.}
This {accumulating experimental evidence} has led to a surge in computational \cite{Oschmann2018} and experimental models \cite{Paniccia2022} aiming to capture the bidirectional interactions between astrocytes and neurons.

Despite this momentum, relatively little effort has been made to analyze these neuron-astrocyte interactions from the perspective of \textit{higher-order interactions} -- that is, interactions that cannot be reduced to the sum of pairwise components \cite{Battiston2021}. 
{Higher-order interactions are increasingly recognized as a key phenomenon underlying the structure and dynamics of complex systems such as the brain \cite{Battiston2021,naturephysicsana2025}. 
From a modeling perspective, higher-order interactions are known to lead to explosive phase transitions and to change the location of the transition points in the parameter space \cite{iacopini2019simplicial, millan2020explosive}, with important consequences for applications e.g. in clinical neuroscience \cite{gatica2022high, digaetano2024neighbourhood, santoro2024higher}.}
In biological networks, higher-order interactions can manifest naturally when a single element such as an astrocyte modulates multiple connections simultaneously, effectively coupling synapses. 
This is particularly relevant in the context of astrocytes, whose broad spatial reach and integrative capacity allow them to coordinate the activity of multiple synapses within their domain \cite{Araque2014}.
Interestingly, many models in the literature already implement mechanisms that can be interpreted as higher-order, albeit implicitly. 
For example, when an astrocyte signal simultaneously alters the release probability or efficacy of two different synapses \cite{Pitta2022}, this constitutes an edge-edge interaction -- a classical case of higher-order modulation in a higher-order network. 
However, such mechanisms are rarely framed or analyzed explicitly through the lens of higher-order network-theory. 
%, which has recently gained traction in the study of complex systems and network science \cite{naturephysicsana2025}.
By making this connection explicit, our study offers a new interpretive lens to understand the computational role of astrocytes. 
We show that, when a single astrocyte regulates multiple synapses within a recurrent circuit, it introduces non-trivial dependencies that shape network-wide behavior -- dependencies that cannot be decomposed into pairwise synaptic dynamics alone. 
This re-framing helps to bridge cellular-level mechanisms and meso-scale network dynamics, suggesting that glial modulation operates as a higher-order control structure within the brain.

Simulations in a minimal recurrent circuit demonstrate that astrocyte-mediated plasticity can effectively prevent the emergence of self-sustained activity (SSA) -- a pathological regime of runaway excitation and loss of responsiveness \cite{Hasselmo1995, Sadeh2021}. 
Crucially, this homeostatic effect only emerges when the astrocyte modulates multiple synapses simultaneously -- i.e., when higher-order interactions are present. 
In this regime, the astrocyte balances excitability and responsiveness by dynamically regulating synaptic efficacy across the circuit.
We have compared low- versus higher-order modulating schemes, demonstrating that, when a single astrocyte modulates several synapses concurrently the regime of adequate-response is maximized. This effect cannot be reproduced by low-order schemes, even when adjusting model parameters or including diffusive coupling between astrocytes.
Furthermore, we have shown that this regime is optimized when the astrocyte modulates synapses internal to the circuit—particularly the recurrent synapse projecting from the read-out neuron back to the read-in neuron. This phenomenon has been validated in larger cyclic networks with N=5 and N=20 neurons, and across different high-order coupling schemes.
These findings support the idea that astrocytes are not merely local modulators, but instead play a system-level regulatory role by coordinating activity across multiple synapses in a structured manner.

Our results align with observations that tripartite synapses are more abundant in brain regions where recurrent connectivity is functionally important—such as the hippocampus and cerebellum \cite{Ventura1999, Tan2021, LeDuigou2014, Sammons2024}. In these regions, it is thought that avoiding runaway excitatory activity and maintaining circuit responsiveness typically requires mechanisms such as inhibitory feedback \cite{Hasselmo1995}. However, based on our findings, alternative or complementary mechanisms involving tripartite synapses and astrocytic modulation may also play an important role.

% A bit of contextualization
Our computational model formalizes this perspective by extending classical short-term plasticity frameworks \cite{Tsodyks1998} to include an astrocyte-mediated facilitation mechanism. 
The resulting tripartite plasticity model generalizes and unifies previous works: when astrocyte modulation is disabled, it reduces to the standard STP model; when pre-synaptic facilitation is removed, it recovers the astrocyte-driven dynamics proposed by Brunel and De Pittà \cite{Pitta2022}. 
This makes the model a flexible platform for studying different regimes of short-time synaptic plasticity, ranging from purely neuronal to glia-integrated mechanisms.

From the higher-order perspective, our model takes advantage of the recently proposed framework of triadic percolation \cite{sun2023dynamic, anapnasnexus}, that studies the dynamic changes in connectivity induced by triadic interactions -- those in which a node regulates the status of the link between two other nodes. 
Recently this framework was applied to the study of axo-axonic interactions in a toy-model neural medium of excitatory neurons \cite{millan2025}, demonstrating that triadic interactions not only lead to dynamic connectivity, but also to complex spatio-temporal patterns of neuronal activity. 
Here we took inspiration from these studies to propose the modeling of neuron-astrocyte networks explicitly as a higher-order network in which astrocytes directly control the status to node-to-node interactions -- the synapses -- and are in turn controlled by neuronal and synaptic activity.

% Limitations
Despite these contributions, our model remains a simplification of biological reality. It focuses on a minimal circuit of excitatory neurons and does not yet incorporate inhibitory dynamics, astrocyte heterogeneity, or spatially distributed networks. 

Moreover, while the HOI perspective offers a compelling conceptual framework, further work is needed to systematically quantify and classify such interactions in biologically realistic networks.
Future studies could expand this framework to larger and more heterogeneous systems, include multiple interacting astrocytes, and explore how astrocyte modulation interfaces with long-term plasticity mechanisms. Additionally, formal tools from higher-order network theory and topological data analysis could be employed to characterize emergent dynamics and identify signatures of glial modulation in experimental data.

This work introduces a computational framework that formalizes astrocyte-mediated modulation of synaptic dynamics as a tripartite plasticity mechanism.
By extending classical short-term plasticity models to include astrocyte-dependent facilitation, our approach captures how a single astrocyte can simultaneously influence multiple synapses, thereby introducing higher-order dependencies into network dynamics. 
This provides a biologically grounded implementation of higher-order interactions in neuronal-{astrocytes} systems.
Overall, this work contributes to a growing body of evidence that astrocytes are key players in neural computation -- not only at the synaptic level but also in the emergent coordination of network-wide activity through higher-order, structured interactions \cite{DePitta2020,Sun2023}.

\section{Acknowledgements}
This work has been supported by Grant No. PID2023-149174NB-I00 financed by the Spanish Ministry and Agencia Estatal de Investigación MICIU/AEI/10.13039/501100011033 and ERDF funds (European Union) (to A.P.M. and J.J.T.). 
A.P.M. also acknowledges financial support by the 'Ramón y Cajal' program of the Spanish Ministry of Science and Innovation (Grant RYC2021-031241-I). G.M. would like to thank the ``Programa Nacional de Becas de Postgrados en el Exterior “Don Carlos Antonio López” - BECAL'' of the Ministry of Economy and Finance of Paraguay for the financial sponsorship to pursue his doctoral studies in the Physics and Mathematics Program of the University of Granada. We thank Maurizio De Pittà for his valuable feedback on this manuscript, which helped enhance both its clarity and depth through his thoughtful comments and suggestions.

\bibliographystyle{unsrt}
\bibliography{references}

\newpage

\newcommand{\asection}[2]{
\setcounter{section}{#1}
\addtocounter{section}{-1}
\section{#2}
}

\setcounter{figure}{0}

\renewcommand{\thefigure}{\Alph{figure}}

\centerline{\huge Supplementary Information}
\vspace*{0.5cm}
\centerline{\Large Astrocyte-Mediated Higher-Order Control of Synaptic Plasticity}
\vspace*{0.5cm}
\centerline{\large Gustavo Menesse, Ana P. Mill\'an \& Joaquín J. Torres}

\asection{1}{Facilitation mechanism in absence of gliotransmitter release or in absence of pre-synaptic facilitation mechanism.}

If there is no gliotransmitters released in a long time, means that $\Theta([Ca^{+2}]_{a}-Ca_{thr})=0$, or that all resources have been consumed $x_{a}^{as}(t)\approx 0$. Then $\gamma_{ija}^{as}(t)=0$, and the release probability will be,

%If there is no gliotranmitters released, it means $\delta([Ca^{+2}]_{a}-Ca_{thr})=0$. Then $\gamma_{ija}^{as}(t)=0$, and the release probability will be,
\begin{CEquation}
u_{ija}(\gamma_{ija}^{as}=0,\gamma_{ij}^{pre},t) = U_{SE} + (1-U_{SE})\gamma_{ij}^{pre} \:,  \label{Eq_U_noglio}
\end{CEquation}
taking the derivative in time we have, 
\begin{equation}
\frac{du_{ij}(\gamma_{ij}^{pre},t)}{dt} = U_{SE} + (1-U_{SE})\frac{d\gamma_{ij}^{pre}}{dt} \:, \nonumber
\end{equation}

Substituting the dynamics of the pre-synaptic activated fraction $\gamma_{ij}^{pre}$ when $\gamma_{ija}^{as}=0$, gives
\begin{equation}
\frac{du_{ij}(\gamma_{ij}^{pre},t)}{dt} = (1-U_{SE})\left[-\frac{\gamma_{ij}^{pre}}{\tau_{pre}} + U_{SE}\left(1-\gamma_{ij}^{pre}\right)\delta(t-t^{spk}_{i})\right] \:, \nonumber
\end{equation}

From \eqref{Eq_U_noglio}, we can express $\gamma_{ij}^{pre}$ and $(1-\gamma_{ij}^{pre})$ as function of $u_{ij}(t)$ as  
\begin{eqnarray}
\gamma_{ij}^{pre}(t) &=& \frac{u_{ij}(t)-U_{SE}}{1-U_{SE}}   \nonumber \\
1-\gamma_{ij}^{pre}(t) &=& \frac{1-u_{ij}(t)}{1-U_{SE}} \:. \nonumber
\end{eqnarray}
And by substituting this expressions in the $u(t)$ dynamics we obtain,
\begin{eqnarray}
\frac{du_{ij}}{dt} &=&  \frac{U_{SE}-u_{ij}(t)}{\tau_{pre}} + U_{SE}(1-u_{ij}(t))\delta(t-t^{spk}_{i}) \:. \nonumber
\end{eqnarray}
Which is the usual pre-synaptic short-term facilitation model \cite{Tsodyks1998,Markram1996}.

Otherwise, if there is no pre-synaptic spikes in a long time, the pre-synaptic activated fraction $\gamma_{ij}^{pre}=0$, following similar reasoning's as the previous case, yields to
\begin{eqnarray}
\frac{du_{ija}}{dt} &=& \frac{U_{SE}-u_{ija}(t)}{\tau_{as}} + G_{a}(t)(\varepsilon-u_{ija}(t))\Theta([Ca^{+2}]_{a}-Ca_{thr}) \: , \nonumber \end{eqnarray}
%\begin{eqnarray}
%\frac{du_{ija}}{dt} &=& \frac{U_{SE}-u_{ija}(t)}{\tau_{as}} + G_{a}(t)(\varepsilon-u_{ija}(t))\delta([Ca^{+2}]_{a}-Ca_{thr}) \: , \nonumber \end{eqnarray}
%which is equivalent to De Pitta model \cite{Pitta2022}.
which is our equivalent to De Pitta model \cite{Pitta2022}, where the only difference is that we have a continuous release if astrocytic cytosolic calcium concentration is above a release threshold, while De Pitta describe it as a singular event in time.

\pagebreak

\section{Neuronal clique activity and astrocytic modulation}

\begin{figure}[ht!]
    \centering
    \includegraphics[width=0.65\linewidth]{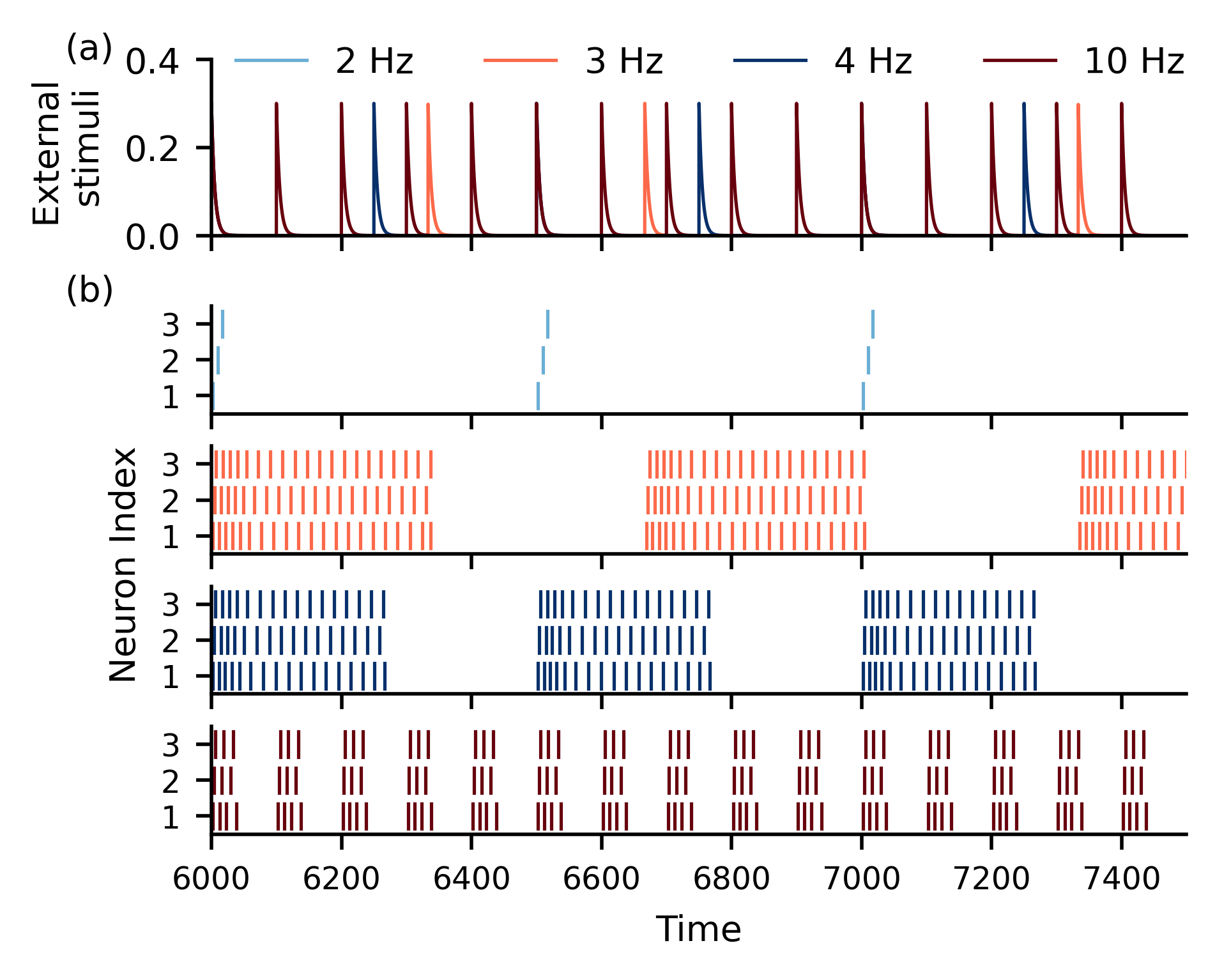}
    \caption{\textbf{Network activity under different stimulation frequency in experimental scheme without gliotransmission}. For different stimulations frequency, if the synaptic current is low, the system shows different responses such as synfire activity (stimuli at 2 Hz), Up-Down states switching via external inputs (3 and 4 hz), short bursting activity ($> 5$ Hz). For this richness to be possible, the synaptic current should be large enough to propagate activity, but small enough to avoid run-away excitatory activity. Here, we use fraction of neurotransmitter in the synaptic cleft $\alpha = 0.5$, neurotransmitter steady-state release probablity $U_{SE}=0.1$ and no gliotransmitter release ($U^{as}_{SE}=0$).}
    \label{fig:Switching}
\end{figure}

\begin{figure}[ht!]
    \centering
    \includegraphics[width=0.9\linewidth]{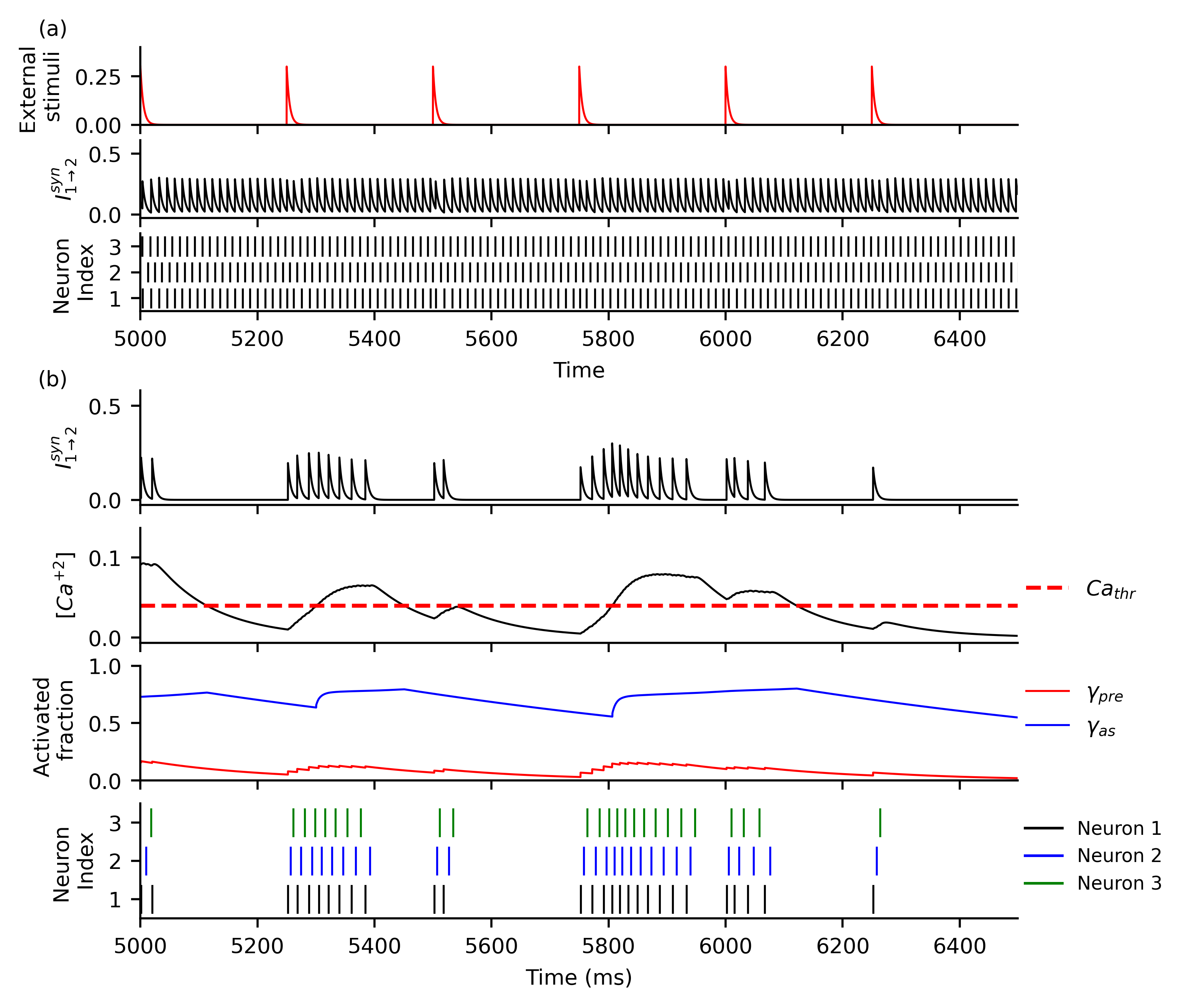}
    \caption{\textbf{Gliotransmission avoid run-away activity caused by an excessive synaptic current}. \textbf{(a)} Run-away activity caused by excess of synaptic current. (a) Top panel illustrates synaptic current from external stimuli, (a) Middle panel shows synaptic current from neuron 1 to 2 and (a) Bottom panel depicts the neuronal spiking raster plot. Increasing the amount of neuroreceptors in the synaptic cleft ($\alpha=0.8$) make the system to enter a self-sustained activity regime where the system becomes practically unresponsive to inputs. \textbf{(b)} In presence of gliotransmission, the triadic interaction modulates the synaptic facilitation, protecting the circuit against run-away excitatory activity. (b) Top panels show synaptic current $I^{syn}_{1\rightarrow 2}$ from neuron 1 to 2 and global calcium concentration $[Ca^{+2}]$ in astrocite. (b) Bottom panels illustrate facilitation activation fractions from both pre-synaptic and astrocitic mechanism ($\gamma_{pre}$ and $\gamma_{as}$) and neuronal spiking raster plot. In both cases, stimulation frequency is 4 Hz, $U_{SE}=0.1$ and $\epsilon=0.01$. The gliotransmission probability used was $U_{S}=0$ in (a) and  $U_{S}=0.1$ in (b). All other parameters are shown in Table \ref{tab:Parameters}.}
    \label{fig:AvoidRunAway}
\end{figure}

\clearpage
\pagebreak

\section{Safe regime vs self-sustained activity (SSA) regime.}

\begin{figure}[ht!]
    \centering
    \includegraphics[width=0.9\linewidth]{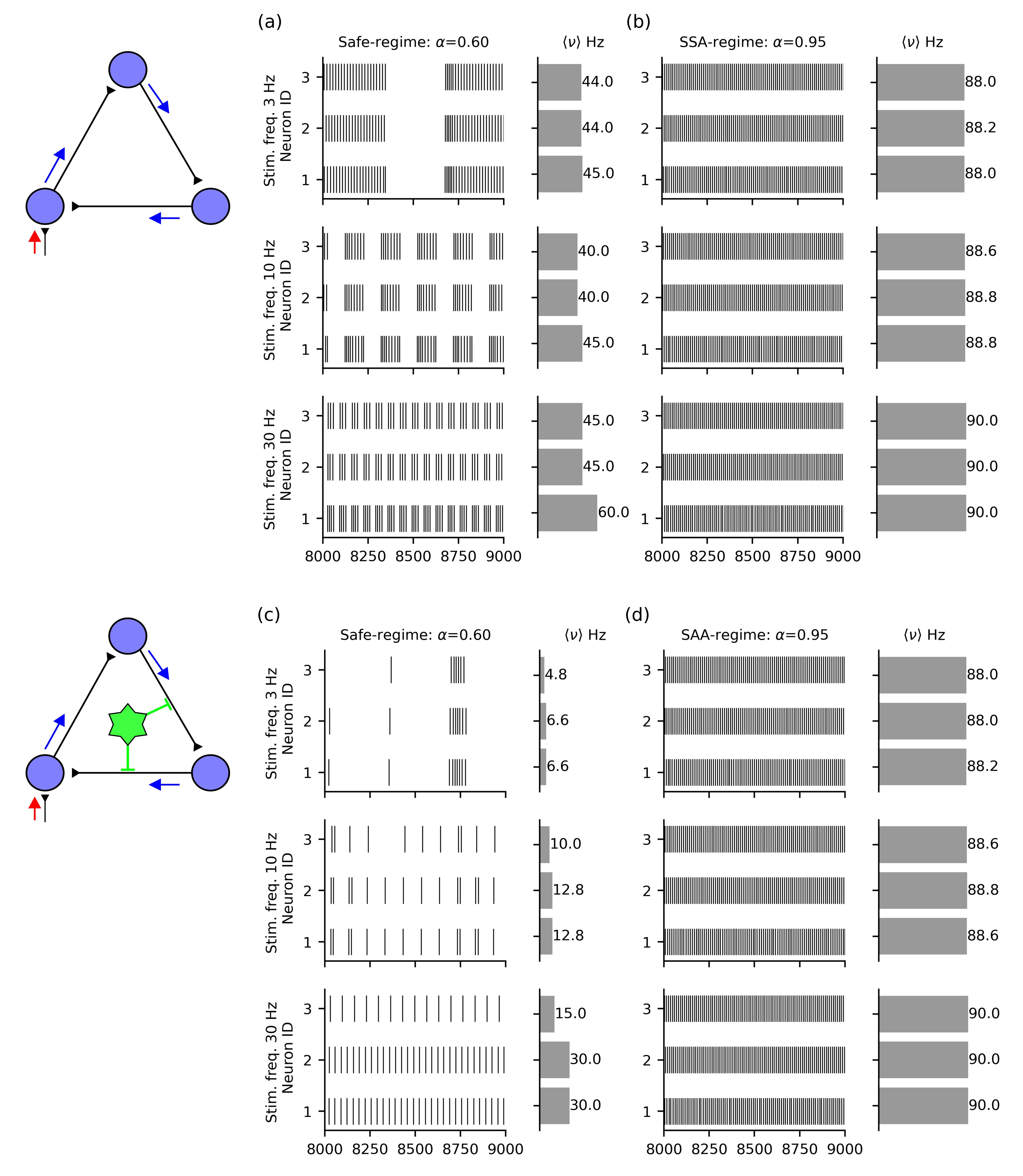}
    \caption{\textbf{Raster plots and average firing rates in safe vs. self-sustained activity regimes under different stimulus frequencies.}  Each panel shows the spike raster (left) and average firing rates (right) for all neurons.  (a) In the safe regime ($\alpha = 0.6$), without astrocyte modulation, the circuit exhibits up–down state switching driven by external input (as in Figure~\ref{fig:Switching}). Neurons show weak, irregular average firing rate responses to stimulus frequency.
(b-d) In the SSA regime, the circuits becomes insensitive to external input, maintaining high firing rates with minimal variation. (c) In the safe regime, with astrocytic modulation, the circuit exhibits strong, frequency-dependent responses, with average firing rates increasing monotonically with stimulus frequency.}
    \label{fig:Safe_vs_SSA}
\end{figure}

\pagebreak
\newpage

\section{Minimum amplitude of synaptic current to propagate activity}

To identify the SSA regime more accurately, we will derive some expressions for the minimal amplitude of synaptic current that, given a specific time and value of membrane potential of the post-synaptic neuron, will generate a spike. If all synaptic current amplitude measured in a given simulation are above this value, means that even if external stimuli stops, the system will maintain activity in a self-sustained manner.

The membrane potential and the synaptic current dynamics are described as,

\begin{subequations}
\begin{align}
\tau_{v}\frac{dV(t)}{dt} &=  -V(t) + R A_{se} y(t) \label{eqn:line-1}\tag{S.2} \\
\frac{dy(t)}{dt} &= -\frac{1}{\tau_{in}}y(t) + \alpha x U \delta(t^{spk}-t)\label{eqn:line-2} \tag{S.3}
\end{align}
\label{eqn:all-lines}
\end{subequations}

%\begin{equation}
%\begin{array}{ccc}
%\tau_{v}\frac{dV(t)}{dt} &=&  -V(t) + R A_{se} y(t)  \tag{S.2}\\
%\frac{dy(t)}{dt} &=& -\frac{1}{\tau_{in}}y(t) + \alpha x U \delta(t^{spk}-t)\tag{S.3} 
%\end{array}
%\end{equation}

Where $R$ is the membrane resistence, $C$ is the membrane capacitance and $\tau_{v}=RC$. $A_{se}$ is the maximum current amplitude, $y(t)$ is the fraction of activated neurotransmitters, $\alpha$ is the fraction of neurotransmitters that stay in the synaptic cleft, $x U$ is the synaptic efficacy and $\tau_{in}$ is the synaptic current time scale.

After a spike at time $t^{spk}=0$ in the pre-synaptic neuron, the fraction of activated neurotransmitter follows

$$y(t) = \left(\alpha x U + y(0)\right)e^{-\frac{t}{\tau_{in}}}$$

Therefore, the change in membrane potential after this event can be computed,

$$\frac{dV(t)}{dt} = -\frac{1}{\tau_{v}}V(t) + \frac{R A_{se}}{\tau_v} \left[\left(\alpha x U + y(0)\right)e^{-\frac{t}{\tau_{in}}}\right]$$

Solving, we obtain,

$$V(t) = \frac{R A_{se}\left(\alpha x U+y(0)\right) \tau_{in}}{\tau_{in}-\tau_{v}} \left[ e^{-\frac{t}{\tau_{in}}} - e^{-\frac{t}{\tau_{v}}} \right] + V(0)e^{-\frac{t}{\tau_v}}$$

For the synaptic current to be capable of generating a spike in the neuron, the maximum of $V(t)$ should be equal or greater than the threshold $V_{thr}$.

The time were the membrane potential will achieve it's maximum value is easily computed finding the extreme of the last equation. This time is,

$$ t_{max} = \frac{\tau_v\tau_{in}}{\tau_{in}-\tau_v} \ln\left\lbrace \frac{1}{\tau_v}\left[\tau_{in} - \frac{V(0)(\tau_{in}-\tau_{v})}{R A_{se}\left(\alpha x U + y(0)\right)}\right]\right\rbrace $$

To simplify notation we use $C = \frac{1}{\tau_v}\left[\tau_{in} - \frac{V(0)(\tau_{in}-\tau_{v})}{R A_{se}\left(\alpha x U + y(0)\right)}\right]$. 

The maximum value of membrane potential is,

$$V_{max} = \left(\frac{\tau_{in}}{\tau_{v}-\tau_{in}}\right)R A_{se}\left(\alpha x U + y(0)\right) \left[C^{\frac{\tau_{v}}{\tau_{v}-\tau_{in}}} - C^{\frac{\tau_{in}}{\tau_v-\tau_{in}}} \right] + V(0)C^{\frac{\tau_{in}}{\tau_{v}-\tau_{in}}}$$

The synaptic current will generate a spike only if $V_{max} \geq V_{thr}$, therefore given a $y(0)$ and $V(0)$, a spike will be generate if the following inequality holds,

\begin{equation}
\left(\frac{\tau_{in}}{\tau_{v}-\tau_{in}}\right)R A_{se}\left(\alpha x U + y(0)\right) \left[C^{\frac{\tau_{v}}{\tau_{v}-\tau_{in}}} - C^{\frac{\tau_{in}}{-\tau_{in}-\tau_v}} \right] + V(0)C^{\frac{\tau_{in}}{\tau_{v}-\tau_{in}}} \geq V_{thr}  \label{Eq_MinCur}\tag{S.4}
\end{equation}
Importantly, if the spike occurs while the post-synaptic neuron is in the refractory period, we need to compute the synaptic current $y(t)$ at instant t when the post-synaptic neuron exits the refractory period. 

In the case of rest membrane value $V(0)=0$ and pre-spike synaptic current $y(0)=0$ the expression simplifies to,

$$\alpha x U \underbrace{\left(\frac{R A_{se}\tau_{in}}{\tau_{v}-\tau_{in}}\right)\left[ \left(\frac{\tau_{in}}{\tau_{v}}\right)^{\frac{\tau_{v}}{\tau_{in}-\tau_{v}}} - \left(\frac{\tau_{in}}{\tau_{v}}\right)^{\frac{\tau_{in}}{\tau_{in}-\tau_{v}}} \right]}_{B} - V_{thr} \geq 0 $$

$$ \alpha x U \geq \frac{V_{thr}}{B} $$

Therefore, we can compute the minimum value of activated fraction amplitude $\alpha x U$ to produce a spike. If astrocytic modulation reduce the release probability $U$, then a larger value of $\alpha$ will be needed to produce a transition to SSA regime.

\subsection*{IV.1 Calculation of SSA transition line}

From the previous analysis, we know how it should be the synaptic current arriving to a post-synaptic neuron to induce a spike. If all synaptic currents in the system satisfy the condition given in Eq.~\ref{Eq_MinCur}, then every emitted spike will reliably trigger a spike in a postsynaptic partner. In this case, each spike induces another, perpetuating activity throughout the network and leading to a self-sustained activity (SSA) regime.

To compute the transition line shown in Figure 2 of the main text, we recorded, during simulation, the exact arrival time of spikes at each neuron, along with the corresponding state of the activated fraction $y(t)$, the membrane potential $V(t)$ at the time of spike arrival ($t=0$), and the values of the plasticity variables $U$ and $x$. We then evaluated whether the inequality in Eq.~\ref{Eq_MinCur} was satisfied for all spike events in the simulation. If this condition holds for all recorded spikes, it indicates that network activity can be sustained even after external stimulation is removed—thus, the system is in the SSA regime.

We validated this criterion by running simulations with $\alpha$ values slightly above and below the computed SSA transition line. After removing the external stimulation, we observed that activity persisted for $\alpha$ above the line and ceased for $\alpha$ below it (not shown), in agreement with the predicted SSA transition.

\pagebreak

%\section{Calcium integration and astrocyte modulation in different modulation schemes.}

%\begin{turnpage}

\section{Calcium integration and astrocyte modulation in different modulation schemes.}
  
\begin{figure}[ht!]
    \centering
    \includegraphics[width=0.95\linewidth]{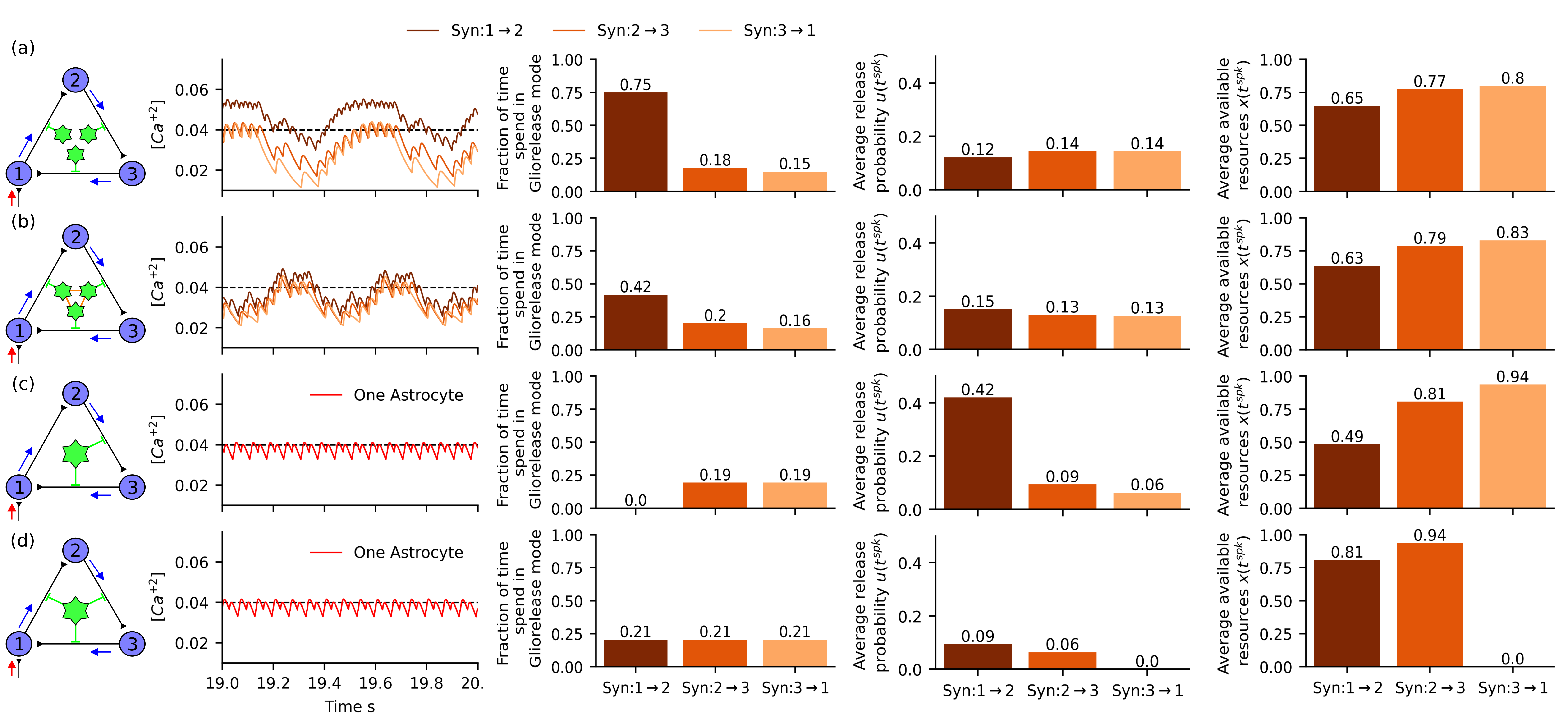}
\caption{\textbf{Calcium dynamics and astrocyte modulation across interaction schemes.}  
Each column shows: calcium concentration over time, gliorelease duration ($[Ca^{2+}] > Ca_{thr}$), average release probability, and available neurotransmitter resources at spike onset ($t^{spk}$). Rows correspond to modulation schemes:  
(a) Low-order via independent astrocytes;  
(b) Low-order with calcium diffusion ($D_{Ca} = 10^{-2}$);  
(c) High-order targeting internal synapses;  
(d) High-order including the external synapse ($1 \rightarrow 2$).  
In (a–b), astrocytes show large calcium fluctuations. The external synapse remains in gliorelease mode over 40\% of the time, while others are active for less than half that. Diffusion in (b) reduces variability but has minimal effect on average modulation.  
High-order schemes (c–d) reduce calcium fluctuations via IP$_3$ integration in a single astrocyte, producing fewer but more effective gliorelease events. In (d), adding the external synapse increases gliorelease time slightly but disrupts activity propagation, silencing the read-out neuron (neuron 3). Stimulus firing rate equal to 30 Hz in all cases.}
    \label{fig:Calcium_dif_schemes}
\end{figure}
%\end{turnpage}
%\clearpage % End the landscape section
%\global\pdfpageattr\expandafter{\the\pdfpageattr/Rotate 90}

\pagebreak

\section{Clique size effect on activity}

\begin{figure}[h]
    \centering
    \includegraphics[width=0.8\linewidth]{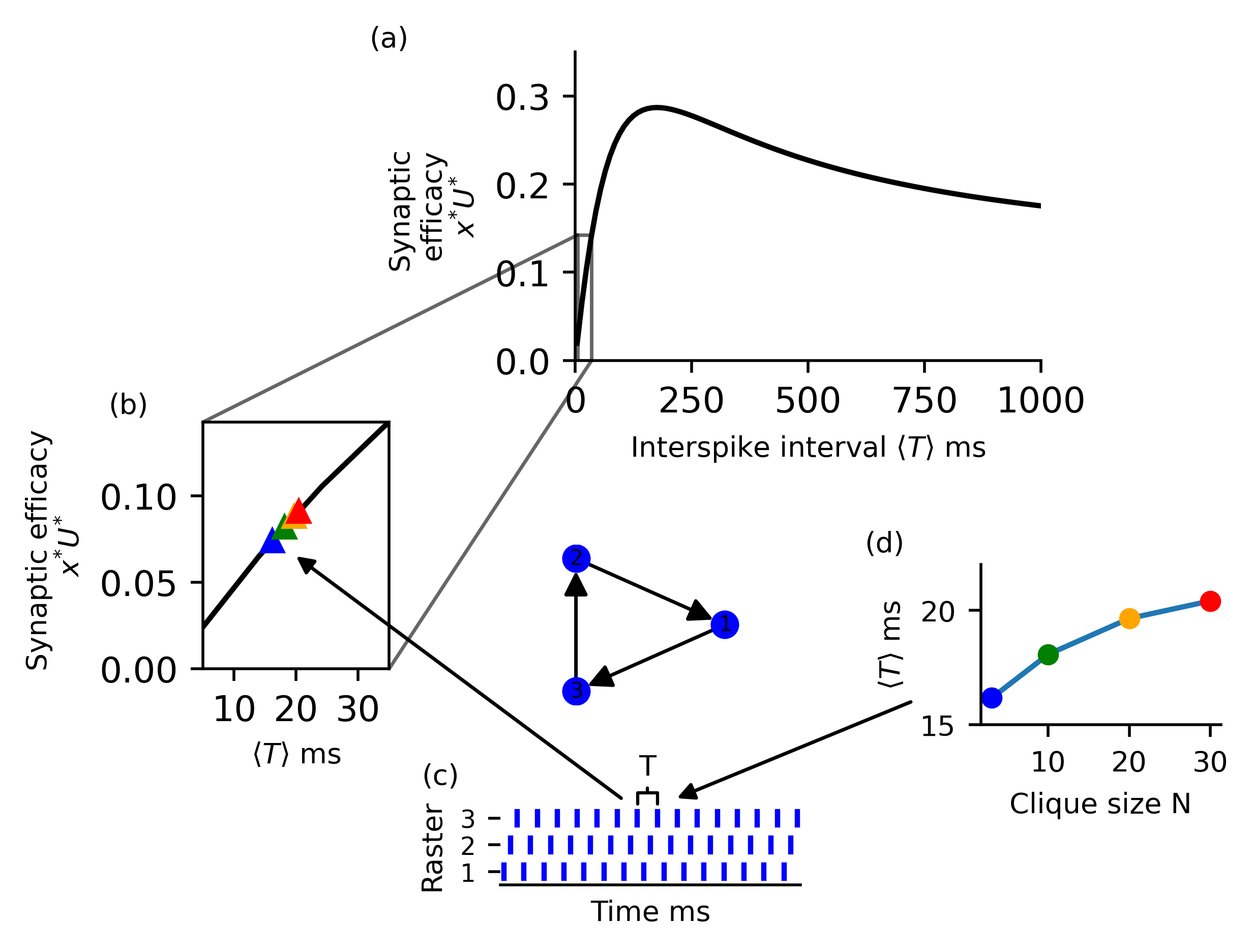}
    \caption{\textbf{Synaptic efficacy as a function of clique size}. In the absence of gliotransmission, (a) the steady-state synaptic efficacy as a function of the pre-synaptic interspike interval can be computed analytically (black solid curve). (b) Synaptic efficacy increases with interspike interval—that is, it decreases with firing frequency. (c) Spike train (raster plot), when the neuronal clique is driven by recurrent activity, the interspike interval corresponds to the time T required for activity to propagate through the entire loop, completing one full cycle. (d) As the size of the clique increases, this propagation time also increases (blue solid curve in the inset), leading to longer interspike intervals. As a result, synaptic efficacy rises with clique size, enhancing recurrent excitation (increase synaptic current) and increasing the vulnerability of the network to enter a self-sustained activity (SSA) regime. Marker's color code correspond to data for different clique size N.}
    \label{fig:RunAwaygetWorst}
\end{figure}

\end{document}